\pdfoutput=1
\documentclass[letterpaper]{article} %
\usepackage{aaai25}  %
\usepackage{times}  %
\usepackage{helvet}  %
\usepackage{courier}  %
\usepackage[hyphens]{url}  %
\usepackage{graphicx} %
\usepackage{natbib}  %
\usepackage{caption} %
\usepackage{algorithm}
\usepackage{algorithmic}
\usepackage{placeins}

\usepackage{newfloat}
\usepackage{listings}
\DeclareCaptionStyle{ruled}{labelfont=normalfont,labelsep=colon,strut=off} %
\floatstyle{ruled}
\newfloat{listing}{tb}{lst}{}
\floatname{listing}{Listing}
\usepackage{xcolor}

\usepackage{xcolor}
\newcommand{\kiran}[1]{}
\newcommand{\shreyasi}[1]{}
\newcommand{\rev}[1]{#1} %

\title{How People Use ChatGPT: Conversation-Level Evidence from India, Nigeria, Brazil, and Pakistan}
\author{
    Shreyasi Roy Chowdhury\textsuperscript{\rm 1},
    Kiran Garimella\textsuperscript{\rm 2}
}
\affiliations{
    \textsuperscript{\rm 1}Independent Researcher, Kolkata, India\\
    \textsuperscript{\rm 2}Rutgers University, New Brunswick, NJ, USA\\
    shreyasi.rc12@gmail.com, kiran.garimella@rutgers.edu
}
\nocopyright

\usepackage{bibentry}

\begin{document}

\maketitle

\begin{abstract}
Public understanding of how people use \rev{LLM-based conversational AI assistants} comes primarily from aggregate platform reports by OpenAI \rev{(for ChatGPT)} and Anthropic \rev{(for Claude)}, which apply fixed taxonomies and inferred demographics to hundreds of millions of users and release only summary statistics that outside researchers cannot re-analyze. We provide a complementary, conversation-level view: complete ChatGPT exports comprising 202,590 conversations from 1,252 users across India, Nigeria, Brazil, and Pakistan, paired with self-reported age and gender and spanning December 2022--February 2026. To our knowledge this is the first conversation-level, demographically-grounded comparison of ChatGPT use across multiple non-Western countries. We use this dataset as a first exploration of three dimensions of usage that each correct a blind spot in the aggregate picture.
\rev{We ask what these users use ChatGPT for (purpose), what they talk about (topics), and how they engage with it (mode of interaction). We answer each question with the platform's own classifiers, which keep our numbers comparable with published aggregates, with unsupervised topic discovery for what those classifiers miss,
and with a thematic analysis of expressive conversations. Personal use accounts for 55--64\% of conversations in every country and coursework is about as common as work, so workplace productivity describes a minority of use. Unsupervised topic discovery surfaces country-specific uses that the OpenAI taxonomy folds into generic categories: health and wellness in India and Brazil, Urdu--English translation in Pakistan, current affairs in Nigeria,
religious questions in Nigeria and Pakistan, and self-reflection in Brazil. Over three years, the share of conversations that seek information declined only modestly, and the share that delegate a task did not grow, while conversations in which users express themselves rose from a few percent to roughly a fifth or more in every country.}
\rev{The same product is thus attached to different local needs in each country, and understanding what adoption means requires conversation-level, country-sensitive measurement alongside global aggregates.}

\end{abstract}

\begin{links}
\link{Dataset}{https://anonymous.4open.science/r/how-people-use-chatgpt-data-1335/README.md}
\end{links}

\section{Introduction}

ChatGPT is now one of the most rapidly adopted technologies in history, with \rev{about 700 million weekly active users by mid-2025}~\citep{chatterji2025chatgpt}. \rev{What those users actually do with it is known almost entirely through the aggregate reports of the company that operates it.} The pace of adoption alone is striking, but the more important observation is that the medium itself is new. The previous generation of widely-used information tools (search engines, online encyclopaedias, the web at large) supports essentially one mode of interaction: a user issues a query and consumes the answer. Conversational AI is the first technology at population scale that supports something different. Users can disclose personal context, frustrations, or emotions, iterate with the model across multiple turns, and interleave self-disclosure with information-seeking and task delegation in the same exchange. This is a qualitatively different form of human-computer interaction from search. \rev{Deliberate use of a chatbot is one of several ways people now encounter AI, alongside AI-generated summaries embedded in search engines and messaging apps. This paper is about the former.}

Understanding what people actually do with this new medium is now a central empirical question for AI research, economics, and public policy, and the stakes are concrete: economists are projecting which occupations will be reshaped by LLMs~\citep{eloundou2023gpts}, governments are debating regulation, and AI labs are pricing tiers and shaping defaults that determine what billions of users see. \rev{The question is sharpest where adoption is growing fastest, and that is where the least is known: OpenAI reports that its fastest growth is now in lower- and middle-income countries~\citep{chatterji2025chatgpt}.}

\rev{Describing that use is harder than counting what people ask about, for three reasons.} The economics literature on LLMs has focused overwhelmingly on workplace productivity~\citep{eloundou2023gpts, brynjolfsson2023generative, noy2023experimental, dellacqua2023navigating}, but health information, translation, educational support, gig-economy navigation, and emotional reflection are forms of value that do not appear in standard labor-market accounting, and we know almost nothing about their prevalence at the conversation level. \rev{The manner of use matters as much as the subject:} because conversational AI uniquely supports back-and-forth, expressive interaction, a complete picture of usage has to account for the new kinds of conversations users have, not just the topics they ask about. \rev{And a taxonomy broad enough to stay stable across hundreds of millions of conversations is necessarily too coarse to show what its categories contain in a given country.}

\rev{Neither of the two kinds of evidence we have settles the matter.} The most influential conversation-level analyses come from platform operators in aggregate~\citep{chatterji2025chatgpt, anthropic2026economic, anthropic2025geographic}\rev{, which classify against taxonomies fixed at the platform, infer demographics from names, and release only summary statistics that outside researchers cannot re-analyze}. \rev{The alternative, data donation by consenting users, has so far been applied to} European and U.S.\ populations~\citep{karnam2026bowling, fang2026aiwrapped}. \rev{The result is that there is} no conversation-level, demographically grounded view of ChatGPT use from the Global South\rev{, and no way to look inside the categories the aggregate reports use}.

We address this with a large-scale data donation in markets where conversation-level data has not previously been available. Through \rev{Clickworker}, a crowdsourcing platform, we recruited 1,252 participants across India, Nigeria, Brazil, and Pakistan, each of whom uploaded a complete ChatGPT data export, \rev{after a client-side script removed names and contact details,} and provided self-reported age and gender. The resulting corpus contains 202,590 conversations spanning December 2022 through February 2026. \rev{The sample is a convenience sample, where two thirds of participants are men, the median age ranges from 24 in Pakistan to 32 in Brazil, and all are digitally literate people who do paid online micro-work and were willing to share their history for compensation. Our estimates describe this population rather than national populations, and gig work also shapes what work-related use looks like in the data.}

Each conversation is characterized on three axes: \emph{why} the user is having it (purpose), \emph{what} it is about (topics), and \emph{how} they are engaging with the model (intent). \rev{We ask of each how the answer differs by country, gender, and age, which the self-reported demographics make possible, and organize the paper around the three corresponding research questions.}

(i) \textbf{RQ1 (Purpose, Section~\ref{sec:work}).} \rev{What share of use is work, coursework, and personal, and how does it vary by country, demographic group, and over time? Using the task-purpose classifier of the Anthropic Economic Index \citep{anthropic2026economic}, we find that personal use dominates in every country and coursework is about as common as work, so the workplace-productivity narrative captures only part of what these users do.}

(ii) \textbf{RQ2 (Topics, Section~\ref{sec:topics}).} \rev{What do users talk about, and which locally prevalent uses does a fixed global taxonomy fail to distinguish? OpenAI's 24-category taxonomy \citep{chatterji2025chatgpt} keeps our numbers comparable with published aggregates, while an unsupervised pipeline run separately for each country surfaces uses that the taxonomy folds into generic categories: health and wellness, translation, religious questions, online earning, and self-reflection.}

(iii) \rev{\textbf{RQ3 (Manner of use, Section~\ref{sec:ade}).} How do users engage with the model, how has that changed over time and within users, and what do expressive conversations contain? Combining the \textit{Asking}/\textit{Doing}/\textit{Expressing} classifier \citep{chatterji2025chatgpt} with language detection and a reading of expressive conversations, we find that information seeking declined only modestly and task delegation did not grow, while \textit{Expressing}, the mode earlier information tools could not host, grew to roughly a fifth of conversations in every country, within individual users as well as in the aggregate.}

\rev{
These findings have implications beyond description. For measurement, global taxonomies need bottom-up discovery beside them to show what adoption means in a given country. For design and safety, a large and growing share of use is expressive and multilingual, so evaluation that is English-first and treats emotional support and task completion as separate domains will miss the typical conversation in these markets. For the economics of AI, coursework and household uses are where much of the value sits for these users.}

\section{Related Work}

\paragraph{Measuring real-world LLM use at scale.}
The most detailed descriptions of how people use large language models come from the platforms that operate them. \citet{chatterji2025chatgpt} classify a sample of ChatGPT conversations drawn from a base of \rev{roughly 700 million weekly active users}, and introduce the 24-category topic taxonomy and the three-way \textit{Asking}/\textit{Doing}/\textit{Expressing} intent frame that we adopt. OpenAI's India report adds a country-level decomposition with name-inferred demographics~\citep{openai2026indiachatgpt}. \citet{tamkin2024clio} introduce a privacy-preserving bottom-up topic-discovery system for Claude conversations, and the Anthropic Economic Index and its geographic follow-up~\citep{anthropic2026economic, anthropic2025geographic} contribute the work/coursework/personal task-purpose label and document a heavy concentration of use in high-income markets. All three are aggregate, fix their categories at the platform, and do not release the conversations. Public corpora sit at the other end of the trade-off: WildChat~\citep{zhao2024wildchat} and LMSYS-Chat-1M~\citep{zheng2023lmsys} release roughly a million conversations each and \citet{deng2023early} analyse publicly shared ones, but in each case users self-selected into a non-standard interface or into sharing, which skews these samples toward technical users. Data donation, in which consenting users hand over their own platform records~\citep{boeschoten2022framework}, avoids both problems and has recently been applied to ChatGPT by \citet{karnam2026bowling} and \citet{fang2026aiwrapped}. \rev{No donation study so far covers the Global South or pairs conversations with self-reported demographics across countries.}

\paragraph{Work, coursework, and everyday assistance.}
A large literature measures the workplace impact of generative AI. \citet{eloundou2023gpts} estimate occupational exposure from task descriptions. Field experiments document gains in customer support, professional writing, and consulting~\citep{brynjolfsson2023generative, noy2023experimental, dellacqua2023navigating}. \citet{bick2024rapid} chart adoption without observing conversation content. \rev{Value that markets do not price falls outside this frame. Economists treat time spent producing goods and services for one's own household as economic activity~\citep{becker1965theory}, free digital goods generate consumer surplus that GDP does not record~\citep{brynjolfsson2019massive}, and the digital-labour literature makes a related point about who produces value online and who is paid for it~\citep{terranova2000free}. \citet{gray2019ghost} describe the crowd-work population from which our participants are drawn.} Education is the largest such domain, with reviews and a global student survey documenting widespread use for explanation, summarizing, and drafting~\citep{kasneci2023chatgpt, baidoo2023education, ravselj2025higher}. \rev{How much everyday LLM use falls into these non-market categories, for whom, and in which countries, has not been measured at the conversation level.}

\paragraph{Geography, language, and culturally situated AI use.}
\rev{Critical accounts of AI argue that its systems and its benefits are shaped by the social and economic settings in which they are built and used rather than being universal~\citep{crawford2021atlas, fuchs2022digital}, and that linguistic representation and translation quality help determine who benefits from them~\citep{nicholas2023lost}.} Frontier models are correspondingly biased toward English-speaking and Protestant European cultural values~\citep{tao2024cultural} and fail on scenarios involving religion, food, and naming~\citep{naous2024beer}, while divides in infrastructure, literacy, language, and gender persist on the user side~\citep{hilbert2011digital, joshi2020state, gillwald2018after, sambasivan2021everyone}. The country-level reports that exist~\citep{openai2026indiachatgpt, anthropic2025geographic} classify against fixed, Western-anchored taxonomies. \rev{What people in these countries use a general-purpose assistant for, conversation by conversation and in their own languages, remains undocumented.}

\paragraph{Affective, relational and disclosive use of conversational AI.}
\rev{People respond socially to computers~\citep{nass2000machines}, in some settings disclose more to a virtual agent than to a person~\citep{lucas2014its}, and experience emotional and relational effects from disclosing to a chatbot comparable to those of disclosing to a human partner~\citep{ho2018psychological}.} \citet{phang2025investigating} find, across platform-scale analysis and a randomised trial, that affective use is concentrated among a small subset of users but has measurable well-being implications. Users of companion products form lasting relationships with them~\citep{skjuve2021chatbot}, experience their loss as bereavement~\citep{banks2024deletion}, and describe the bond in the vocabulary of human relationships~\citep{rocha2025unbreakable}, anticipating the parasocial concerns raised by~\citet{turkle2017alone}. \rev{This literature studies either products designed for relationships or affect detected at scale. It says little about the more common case in a general-purpose assistant, where a user discloses a health worry or a religious question in the course of asking for something.}

\rev{Across these strands the existing evidence is aggregate, workplace-centred, or bound to platform-defined categories, and it comes almost entirely from high-income countries. We address these gaps with conversation-level, demographically grounded data from four Global South countries.}

\section{Dataset}
\label{sec:dataset}
\subsection{Data Collection}

We recruited participants from four Global South countries (India, Nigeria, Brazil, and Pakistan) through Clickworker, a global crowdsourcing platform. Participants were required to be active ChatGPT users and were asked to export their complete ChatGPT conversation history using OpenAI's built-in data export feature, which generates a comprehensive archive including all conversations, model metadata, and account information. Participants also provided basic demographic information (age and gender) during registration. We created a simple interface for the participants to export and upload their ChatGPT histories. After the user downloads their export from OpenAI and before it is uploaded to our servers, a client-side script runs locally on the participant's machine to strip personal information (names, email addresses, etc.). Participants were compensated for their participation in accordance with the local minimum wage and approved by our IRB. \rev{Appendix~\ref{app:ethics} describes consent, de-identification, storage, and the release policy.} The data collection happened between December 2025--Feb 2026. Our approach parallels concurrent work by \citet{karnam2026bowling}, who collected ChatGPT exports via GDPR data-export rights, and by \citet{fang2026aiwrapped}, who designed a privacy-preserving ``wrapped''-style pipeline for the same purpose. The key difference is our focus on recruited, demographically profiled users across four Global South countries rather than convenience samples in Europe or the United States.

Table~\ref{tab:dataset-overview} summarizes the key characteristics of our dataset. In total, we collected conversation exports from 1,252 users encompassing 202,590 unique conversations, spanning from December 2022 (shortly after ChatGPT's launch on November 30, 2022) through February 2026, a period of over three years covering the introduction of GPT-3.5, GPT-4o, and the GPT-5 family.
India contributes the largest share of users (44.5\%) and conversations (43.9\%), followed by Nigeria (19.4\% of users), Brazil (19.6\%), and Pakistan (16.5\%). The high variance in conversations per user (IQR of 15--193 overall) reflects a mix of casual and power users across all countries, with Nigeria having the highest median engagement (91 conversations per user).

\begin{table*}[t]
  \centering
  \begin{tabular}{lrrrrrr}
    \hline
    \multicolumn{1}{c}{\textbf{Country}} &
    \multicolumn{1}{c}{\textbf{Users}} &
    \multicolumn{1}{c}{\textbf{Conversations}} &
    \multicolumn{1}{c}{\textbf{Mean}} &
    \multicolumn{1}{c}{\textbf{Median}} &
    \multicolumn{1}{c}{\textbf{P25--P75}} &
    \multicolumn{1}{c}{\textbf{Date Range}} \\
    \hline
    India    & 557   & 88,958  & 159.7 & 51  & 11--197  & Dec 2022 -- Feb 2026 \\
    Nigeria  & 243   & 44,114  & 181.5 & 91  & 20--225  & Dec 2022 -- Dec 2025 \\
    Brazil   & 246   & 40,067  & 162.9 & 62  & 26--159  & Dec 2022 -- Dec 2025 \\
    Pakistan & 206   & 29,451  & 143.0 & 55  & 13--153  & Dec 2022 -- Dec 2025 \\
    \hline
    \textbf{Total} & \textbf{1,252} & \textbf{202,590} & \textbf{161.8} & \textbf{60} & \textbf{15--193} & \textbf{Dec 2022 -- Feb 2026} \\
    \hline
  \end{tabular}
  \caption{Dataset overview by country. The last four columns describe the distribution of conversations per user.}
  \label{tab:dataset-overview}
\end{table*}

Table~\ref{tab:demographics} presents the demographic composition of our sample. Overall, 65.9\% of users are male and 34.1\% are female. However, this gender distribution varies substantially across countries: India (77.2\% male) and Pakistan (75.7\% male) show the most skewed ratios, while Brazil is the only country where female users outnumber males (55.3\% female). Nigeria shows near-parity with a slight male majority (53.1\%).
Age distributions also differ markedly: India and Pakistan skew young (median age 25 and 24, respectively, with over 50\% in the 18--25 bracket), while Brazil is notably older (median 32, with 37\% over 36). Nigeria falls in between (median 30). These demographic differences likely reflect both the age profiles of Clickworker participants in each country and broader patterns of technology adoption.

\begin{table*}[t]
  \centering
  \begin{tabular}{l r r r r r r}
    \hline
    & \multicolumn{2}{c}{\textbf{Gender}} & \multicolumn{3}{c}{\textbf{Age Group}} & \\
    \cline{2-3} \cline{4-6}
    \textbf{Country} & \textbf{Male} & \textbf{Female} & \textbf{18--25} & \textbf{26--35} & \textbf{36+} & \textbf{Mean Age} \\
    \hline
    India    & 430 (77.2\%) & 127 (22.8\%) & 287 (51.5\%) & 187 (33.6\%) & 83 (14.9\%)  & 27.8 \\
    Nigeria  & 129 (53.1\%) & 114 (46.9\%) & 56  (23.0\%) & 133 (54.7\%) & 54 (22.2\%)  & 31.0 \\
    Brazil   & 110 (44.7\%) & 136 (55.3\%) & 45  (18.3\%) & 110 (44.7\%) & 91 (37.0\%)  & 33.3 \\
    Pakistan & 156 (75.7\%) & 50  (24.3\%) & 118 (57.3\%) & 55  (26.7\%) & 33 (16.0\%)  & 27.1 \\
    \hline
    \textbf{Total} & \textbf{825 (65.9\%)} & \textbf{427 (34.1\%)} & \textbf{506 (40.4\%)} & \textbf{485 (38.7\%)} & \textbf{261 (20.8\%)} & \textbf{29.4} \\
    \hline
  \end{tabular}
  \caption{Demographic composition by country, gender, and age group (user counts).}
  \label{tab:demographics}
  \vspace{-\baselineskip}
\end{table*}

\noindent\textbf{Nomenclature.} A \emph{conversation} is a single ChatGPT session (one page in the ChatGPT web or mobile UI) and contains a back-and-forth sequence of \emph{messages}. Each message carries the sender role (user, assistant, or tool), text content, and metadata.

\noindent\textbf{Sample bias.} The sample is \emph{not} designed to be demographically representative: it is a convenience sample of Clickworker participants willing to export and share their ChatGPT history for compensation, with no stratification on age, gender, education, or urban/rural residence. Results should be interpreted as patterns \emph{conditional on} being an active, digitally literate, Clickworker-accessible ChatGPT user, not as estimates of country-wide prevalence. Despite these limits, complete user-level ChatGPT exports paired with self-reported demographics are, to our knowledge, not publicly available at this scale for any of these four countries.
\rev{Two further consequences follow from the recruitment channel and the product. Participants do paid micro-work and many are college students. Fewer of them hold salaried office jobs than the general adult population, so the work share we observe is likely lower, and differently composed (resumes, applications, client correspondence, content production), than it would be in an office-worker sample. And this is a study of ChatGPT use, not of AI use: people who reach AI mainly through assistants embedded in WhatsApp or search engines, or through Gemini, do not appear in our data. We study ChatGPT because its built-in export gives users a complete, timestamped history of their conversations, which is what makes a longitudinal donation study possible.}

\section{Methods}
\label{sec:methods}

Every conversation is passed through four classifiers across three analytical axes: (i) OpenAI's published 24-category topic taxonomy~\citep{chatterji2025chatgpt}, (ii) unsupervised topic discovery via BERTopic with Gemini embeddings~\citep{grootendorst2022bertopic}, (iii) the \textit{Asking/Doing/Expressing} intent classifier of~\citet{chatterji2025chatgpt}, and (iv) the \textit{work/coursework/personal} task purpose from the Anthropic Economic Index~\citep{anthropic2026economic}. \rev{Each conversation also receives a detected language.} Trend analyses aggregate at monthly granularity and exclude months with fewer than 20 conversations. \rev{Each research question draws on a specific subset of these measurements: RQ1 on (iv) and the clusters within each purpose, RQ2 on (i) and (ii), and RQ3 on (iii) together with language detection, within-user trends, and a qualitative reading of expressive conversations. Appendix Table~\ref{tab:rq_methods} sets out that mapping alongside how each measurement is validated and where its results appear.} Classifier validation against human gold labels is reported in Appendix~\ref{sec:classifier_validation}.

\rev{Three of our four labels come from the platforms whose reports we set out to complement, and we use them deliberately: the only population-scale descriptions of ChatGPT use are stated in these vocabularies, so classifying donated conversations the same way is what makes a direct comparison possible and what shows where the aggregate picture is incomplete. The cost is twofold. The categories encode a platform's view of what matters (\textit{Practical Guidance} does not distinguish a question about symptoms from one about religious practice), and re-implementing the classifiers with the platform's own models means the measurement instrument shares a vendor with the product being measured. We limit this by validating every classifier against human labels (Appendix~\ref{sec:classifier_validation}) and by pairing the taxonomy with an unsupervised pipeline that is tied to no taxonomy, which is where country-specific uses appear. The choice still shapes what can be known, since a taxonomy built for a global product will recognize the uses its designers anticipated more readily than others, and we return to this in the Limitations.}

\noindent\textbf{(i) OpenAI taxonomy.} \rev{We assign each conversation to one of the 24 topics of \citet{chatterji2025chatgpt}, and to the coarse groups they define, using GPT-5-mini with their published category definitions and classification prompt. To bound cost we label at the conversation level rather than at every user message, using the first five user--assistant turns as context.}

\noindent\textbf{(ii) Unsupervised topic discovery.} The OpenAI taxonomy is deliberately coarse so that it stays stable across hundreds of millions of conversations: a bucket like \textit{Seeking Information} covers anything from symptom interpretation to celebrity trivia to legal clarifications. We therefore complement it with an unsupervised pipeline that is free to discover whatever structure exists in the data. We embed each conversation with Google's \texttt{gemini-embedding-001} over its first ten user--assistant turns, run MiniBatch $K$-means with $k=500$ to obtain narrow micro-clusters, and merge the centroids by agglomerative clustering into roughly 50 topics per country. Labels are drafted by \texttt{gpt-4o-mini}, adjudicated by Claude Sonnet 4.6 acting as a judge, and reviewed by the authors. \rev{We chose this mix for multilingual embedding quality, for cheap summarisation at scale, and so that no model grades its own labels.} The pipeline yields 50, 45, 36, and 53 topics for India, Nigeria, Brazil, and Pakistan respectively, covering 91--95\% of conversations per country, which we aggregate into ten cross-country themes by keyword matching on cluster labels. Cluster-level and theme-level shares therefore live at different granularities, and a theme's share is at least as large as any constituent cluster's. Hyperparameters, dendrograms, the reassignment rule for low-confidence conversations, and the judge prompts are in Appendix~\ref{app:bertopic-detail}.

\noindent\textbf{(iii) Asking, Doing, Expressing.} \citet{chatterji2025chatgpt} introduced conversation-level classifiers that measure \emph{how} a user is using the model, orthogonally to the topic of the conversation, and \citet{anthropic2026economic} adopted the same framework for Claude. We replicate two of them with the GPT-5-mini API and the released prompts. \rev{Our intent definitions depart from theirs in one respect: they treat \textit{Expressing} as whatever is neither a request for information nor a request to perform a task, whereas we define it as sharing emotions, statements, or reflections, and add an \textit{Other} category for the residue. Because intent can vary within a conversation we classify each user message, using the previous five turns as context, and take the majority label across a conversation's user messages. Ties retain multiple labels and are reported as normalised shares.}

\noindent\textbf{(iv) Work, coursework, personal.} \rev{Using the prompt of \citet{anthropic2026economic}, each conversation is labelled \textit{Work} (tasks related to the user's profession), \textit{Coursework} (coursework or academic help), or \textit{Personal} (neither of the other two).}

\noindent\rev{\textbf{Qualitative reading of \textit{Expressing} conversations.}}
\label{sec:expressing-theme}
\rev{To see what \textit{Expressing} conversations contain, one of the authors read 100 conversations per country, sampled at random from those whose majority label was \textit{Expressing}. Each was read in full, including the assistant's turns: directly when in English or Hindi, and in English translation when in Urdu or Brazilian Portuguese. For each conversation the reader noted what the user disclosed and what they asked for, and grouped the recurring patterns into the themes reported in Section~\ref{sec:ade}. This is an inductive qualitative reading informed by thematic analysis \citep{braun2006using} rather than a formal coding study: one reader, no codebook taken from the platform classifiers, and no inter-coder agreement. It characterizes the forms that expressive conversations take and does not estimate their prevalence.}

\noindent\rev{\textbf{Within-user trends.}}
\label{sec:within-user}
\rev{Trends over calendar time mix changes in how the same people use the model with changes in who is using it. To separate them we restrict to users with at least 30 conversations spanning at least 12 months (530 users, contributing 78\% of all classified conversations) and (a) compare intent shares by calendar quarter across first-use cohorts, so that parallel movement across cohorts indicates behavioural rather than compositional change, and (b) compare each user's intent shares in their first three active months with their last three, reporting the mean within-user change with user-clustered bootstrap intervals over 10,000 resamples of users.}

\noindent\rev{\textbf{Statistical testing.}}
\label{sec:stats}
\rev{Demographic attributes are defined at the user level and users contributed unequal numbers of conversations, so we test differences in conversation-level prevalence with a user-clustered bootstrap: users are resampled within each demographic group while retaining all of their conversations, and we report 95\% percentile intervals from 5,000 resamples. Gender analyses compare women and men. Age analyses compare all three pairs among 18--25, 26--35, and 36+. We apply the Benjamini--Hochberg correction \citep{benjaminihochberg1995} within each family of comparisons, a family being the full set of topic (or purpose) by group comparisons within one country and taxonomy, and mark the differences that survive at $q<.05$. The same procedure is applied to the task-purpose shares. Given the observational design and unequal group and cluster sizes, we read the surviving patterns descriptively rather than as population-level or causal effects.}

\noindent\textbf{Language detection.} We detect the language of each conversation with the \texttt{langdetect} library (fixed seed) applied to the concatenated user messages, capped at the first 5{,}000 characters. Conversations with fewer than 20 characters of user text are marked \textit{too short}. \rev{Because \texttt{langdetect} is trained on standard orthographies and is known to misassign romanized Hindi and Urdu and code-mixed text, we keep only a binary label: English or not for India, Nigeria, and Pakistan, and Portuguese or not for Brazil.}

\section{Results}
\label{sec:results}

We organize the results around the three \rev{research questions} introduced above: \emph{purpose} (Section~\ref{sec:work}), \emph{topics} (Section~\ref{sec:topics}), and \emph{intent} (Section~\ref{sec:ade})\rev{, which answer RQ1 to RQ3 respectively.}

\subsection{RQ1: Work, Coursework, and Personal Use}
\label{sec:work}

\begin{figure}[t]
  \centering
  \includegraphics[width=\columnwidth]{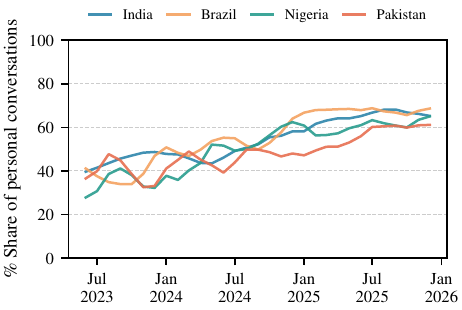}
  \caption{Temporal share of personal conversations relative to work and coursework (3-month rolling average), by country, June 2023 onward.}
  \label{fig:personal_label_trends_line}
	\vspace{-\baselineskip}
\end{figure}

Personal use dominates everywhere: 61.5\% of conversations in India, 63.7\% in Brazil, 55.0\% in Nigeria, and 55.4\% in Pakistan, and the share has been steadily increasing in all the countries (Figure~\ref{fig:personal_label_trends_line}). The 18.5\% work share in our Indian subsample sits well below the 27\% reported globally by \citet{chatterji2025chatgpt}, consistent with the younger composition of our sample and its larger share of students.
\rev{The work/non-work split has been remarkably stable since mid-2024 in India, Nigeria, and Pakistan. Brazil is the exception, having begun in early 2023 with an almost even split before converging to the $\sim$80\% non-work share seen elsewhere (Appendix Figure~\ref{fig:work_label_trends}).}

Appendix Figure~\ref{fig:work-personal} shows the breakdown by age and gender across the four countries. Work and coursework are of roughly comparable prevalence overall across all four countries (20.8\% vs.\ 19.6\%), with the youngest cohort skewing toward coursework and the older cohort toward work (as expected, acting as a sanity check).
\rev{Women's coursework share exceeds men's in India (26\% vs.\ 18\%), Pakistan (33\% vs.\ 21\%), and Nigeria (25\% vs.\ 20\%), but not in Brazil (12\% vs.\ 23\%), where the sample is older and more work-oriented (Appendix Figure~\ref{fig:work-personal}, bottom). This pattern reinforces the female over-representation we see on the \textit{Education/Academic} theme in the topics section below.}
\rev{We tested these gender gaps with the user-clustered bootstrap of Section~\ref{sec:stats}, comparing women and men within each country on the share of conversations in each purpose category. India's coursework gap is the one difference that survives Benjamini--Hochberg correction. The gaps in Pakistan and Nigeria run in the same direction but their confidence intervals include zero (Appendix Table~\ref{tab:work_gender_results}).}

\subsubsection{Topics within work, coursework, and personal conversations}
Under the OpenAI taxonomy, \textit{Writing} and \textit{Technical Help} together account for roughly 45--50\% of work conversations in every country, with \textit{Writing} alone consistently the single largest work bucket (28--42\%) (Appendix Figure~\ref{fig:openai-topic-work}). Our unsupervised BERTopic analysis resolves this into a small number of specific recurring work tasks: resume drafting, job applications and interview preparation, professional email, and content creation for social media platforms (Appendix Figure~\ref{fig:work-topics}). \rev{\textit{Online earning} is a top-5 work cluster in India, Pakistan, and Brazil, and coding assistance is the second most common type of work conversation (Appendix Figure~\ref{fig:work-topics}).}

\begin{figure}[t]
  \centering
  \includegraphics[width=0.5\textwidth]{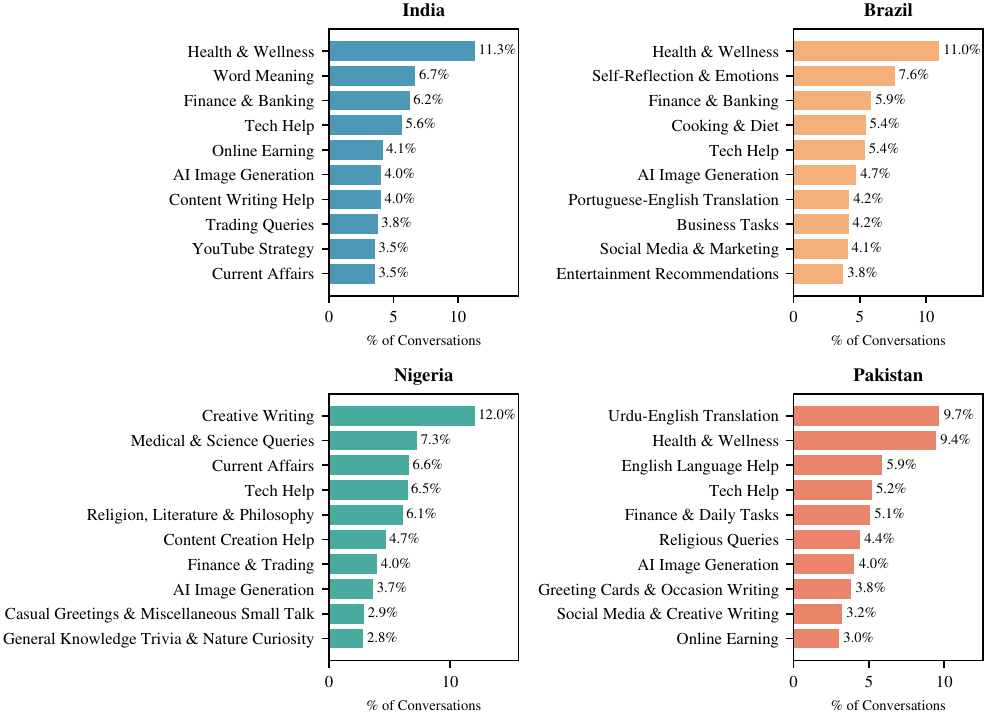}
  \caption{Top 10 unsupervised topic clusters for personal conversations, by country.}
  \label{fig:personal-topics}
\end{figure}

Personal use, by contrast, is anchored everywhere by health and wellness, beyond which it diverges sharply by country, with creative writing and current affairs in Nigeria, Urdu--English translation and religious queries in Pakistan, and self-reflection in Brazil (Figure~\ref{fig:personal-topics}). Coursework conversations are dominated by STEM subjects in every country.

\subsection{RQ2: Topics}
\label{sec:topics}

We characterize topical content through two complementary lenses: the supervised OpenAI 24-category taxonomy~\cite{chatterji2025chatgpt}, \rev{which allows comparison with published aggregates}, and an unsupervised clustering pipeline for topic discovery.

\subsubsection{OpenAI taxonomy}
The supervised picture is close to OpenAI's reported global breakdown (Practical Guidance dominates at 28.8\%, followed by Seeking Information at 22.0\%, Writing at 21.5\%, and Technical Help at 8.5\%), though our sample over-indexes mildly on information-seeking and writing and under-indexes on technical help.
India leads in computer programming (7.6\%) and tutoring/teaching, consistent with the OpenAI Signals finding that Indian users are roughly three times above the global median for coding queries~\cite{openai2026indiachatgpt}. Nigeria has the highest rate of personal writing and communication. Brazil has the highest health and self-care share and the lowest programming share, and Pakistan leads in translation.

Temporally, the early period of ChatGPT adoption was dominated by writing and technical help (most dramatically in Brazil, where \textit{Writing} briefly exceeded 40\% before collapsing), while by 2025 practical guidance and seeking information had become the dominant categories, a convergence that mirrors the late-2025 snapshot OpenAI reports for India~\cite{openai2026indiachatgpt} (Appendix Figure~\ref{fig:topic-trends-openai-lines}).

Gender and age-conditional patterns broadly reproduce, at the individual-conversation level, the name-inferred patterns OpenAI reports for India~\cite{openai2026indiachatgpt}: men over-index on technical help, women on practical guidance and self-expression (Appendix Figure~\ref{fig:gender-topics}).
\rev{Several gender differences have bootstrap intervals that exclude zero, but none survives Benjamini--Hochberg correction (Appendix Table~\ref{tab:taxonomy_gender_results}):
In India, women show higher prevalence of \textit{tutoring or teaching}.  In Brazil, \textit{computer programming} is more prevalent among men, whereas \textit{personal writing or communication} is more prevalent among women. In Nigeria, women show higher prevalence of \textit{health, fitness, beauty or self-care} and \textit{cooking and recipes}. In Pakistan, men show higher prevalence of \textit{purchasable products} and \textit{relationships and personal reflection}. Given uneven gender-group sizes and substantial variation in conversations per user, the remaining visible differences are interpreted as descriptive patterns within this sample rather than broader population-level generalisations.}

\begin{figure}[t]
  \centering
  \includegraphics[width=0.5\textwidth]{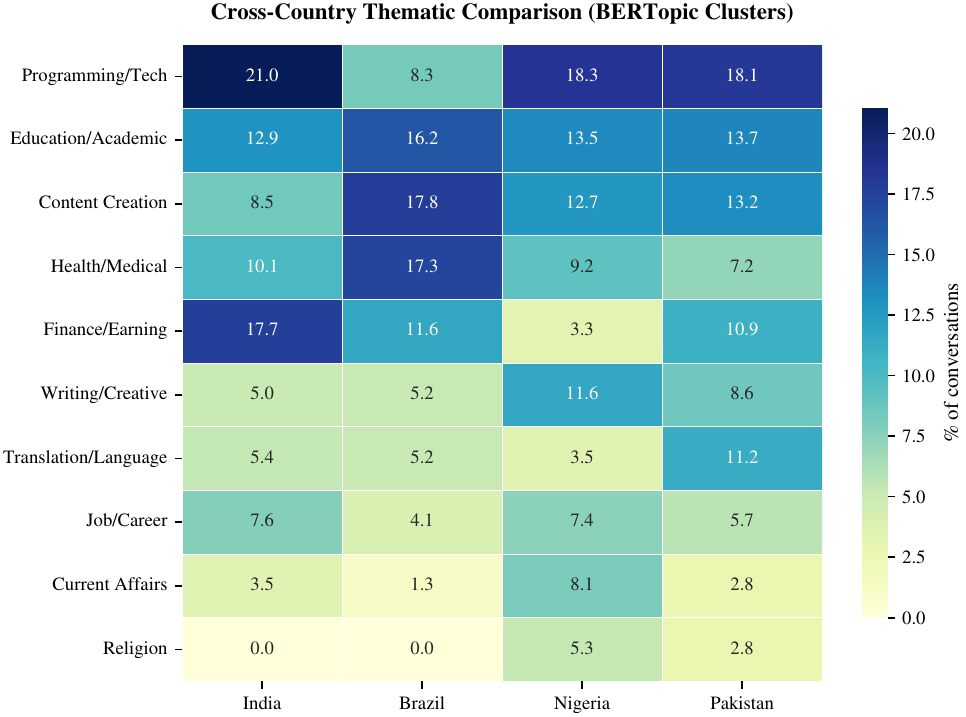}
  \caption{Cross-country distribution of unsupervised topic clusters, rolled up into ten broad themes (share of each country's conversations).}
  \label{fig:theme_heatmap}
\end{figure}

\begin{figure*}[t]
\centering
\includegraphics[width=0.75\textwidth]{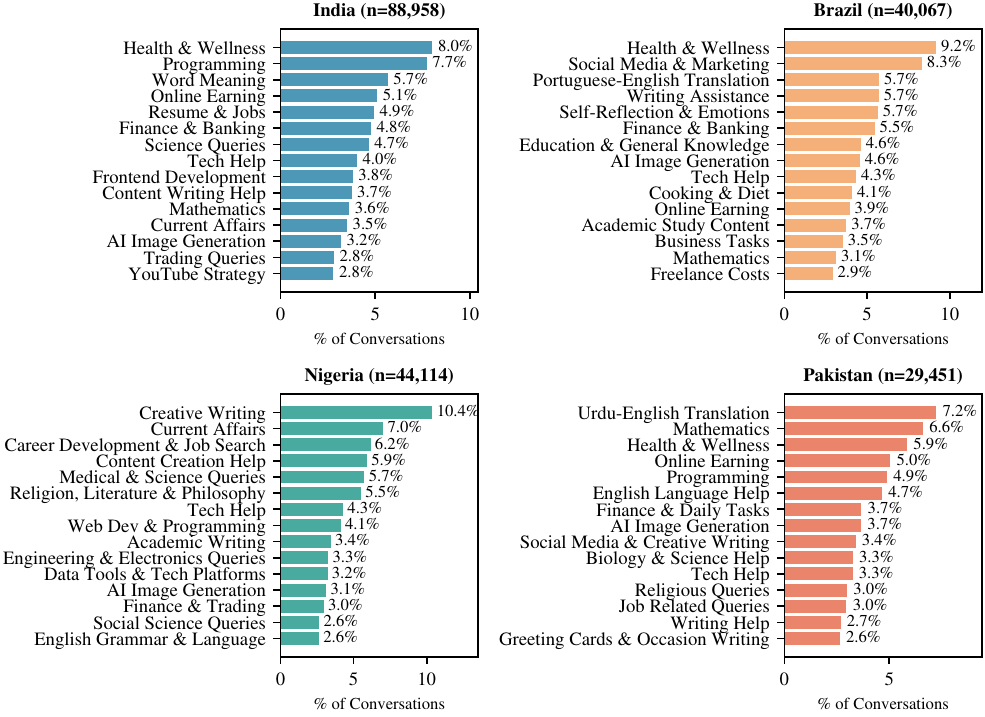}
\caption{Top 15 unsupervised topic clusters per country.}
\label{fig:unsup_topics}
\end{figure*}

\subsubsection{Topic Discovery through Unsupervised Clustering}

The more revealing view comes from the unsupervised pipeline, which is free to discover whatever structure the data contains and so surfaces use cases the fixed taxonomy compresses into generic buckets. Figures~\ref{fig:unsup_topics} and~\ref{fig:theme_heatmap} show the results. \textit{Health and wellness} is the largest cluster in India (8.0\%) and Brazil (9.2\%) and a top-5 cluster in Pakistan\rev{. At the theme level it accounts for 10.1\% of Indian and 17.3\% of Brazilian conversations}.
This is consistent with~\citet{karnam2026bowling}, who document substantial growth in health and mental-health usage over time.
\rev{Several clusters have no counterpart in the OpenAI taxonomy. Urdu--English translation is the single largest cluster in Pakistan (7.2\%). Self-reflection and emotional conversations form a top-5 cluster in Brazil (5.5\%) and appear in no other country's top 15. Online earning is a top-5 cluster in India (5.7\%) and Pakistan (5.9\%), and current affairs is the second-largest cluster in Nigeria (7.0\%).} Aggregating the per-country clusters into ten cross-country themes (Figure~\ref{fig:theme_heatmap}) makes the differences visible at a glance. \textit{Finance/Earning} concentrates in India (17.7\%) and Brazil (11.6\%) but is nearly absent in Nigeria (3.3\%). \textit{Programming/Tech} is most prominent in India and least in Brazil. \textit{Religion} appears exclusively in Nigeria and Pakistan, and \textit{Translation/Language} is highest in Pakistan.

Gender and age gradients under the unsupervised lens are sharper than under the taxonomy because the clusters are thematically tighter. \rev{The differences whose bootstrap intervals exclude zero (Appendix Table~\ref{tab:demographic_topic_results}) run in consistent directions: \textit{Health/Medical} and \textit{Job/Career} lean female (Nigeria, Brazil), while \textit{Programming/Tech}, \textit{Finance/Earning}, \textit{Current Affairs}, and \textit{Religion} lean male (Brazil, Nigeria, Pakistan; Appendix Figure~\ref{fig:gender_bertopic}). Only the Brazil \textit{Programming/Tech} difference remains significant after FDR correction, and we therefore interpret the remaining patterns as descriptive trends within this sample.}

The age picture shows a broadly similar generational gradient across the four countries: 18--25 year-olds concentrate more on education and programming, while the 36+ cohort shifts toward finance/career and civic topics (Appendix Figure~\ref{fig:age_bertopic}). \rev{User-clustered bootstrap analyses provide support for several age-related differences, including higher prevalence of \textit{Education/Academic} among younger users in India, Brazil and Nigeria, substantially higher prevalence of \textit{Programming/Tech} among younger users in India and Pakistan, and \textit{Finance/Earning} among older users in India (Appendix Table~\ref{tab:demographic_topic_results}). After FDR correction, the clearest age differences are the higher prevalence of \textit{Education/Academic} among 18--25 than 36+ users in Brazil, and of \textit{Programming/Tech} among 18--25 than 36+ users in Pakistan. Given the uneven demographic group sizes and variation in the number of conversations per user, we interpret these patterns descriptively and as trends within this sample rather than as broader population-level generalisations.}

\subsection{RQ3: Asking, Doing, Expressing}
\label{sec:ade}

We apply the three-way \textit{Asking}/\textit{Doing}/\textit{Expressing} intent classifier of \citet{chatterji2025chatgpt} (\textit{Asking} = seeking information or decision support; \textit{Doing} = requesting the model to execute a task; \textit{Expressing} = using the model for reflection or emotional communication).

\begin{figure}[t]
    \centering
    \includegraphics[width=0.5\textwidth]{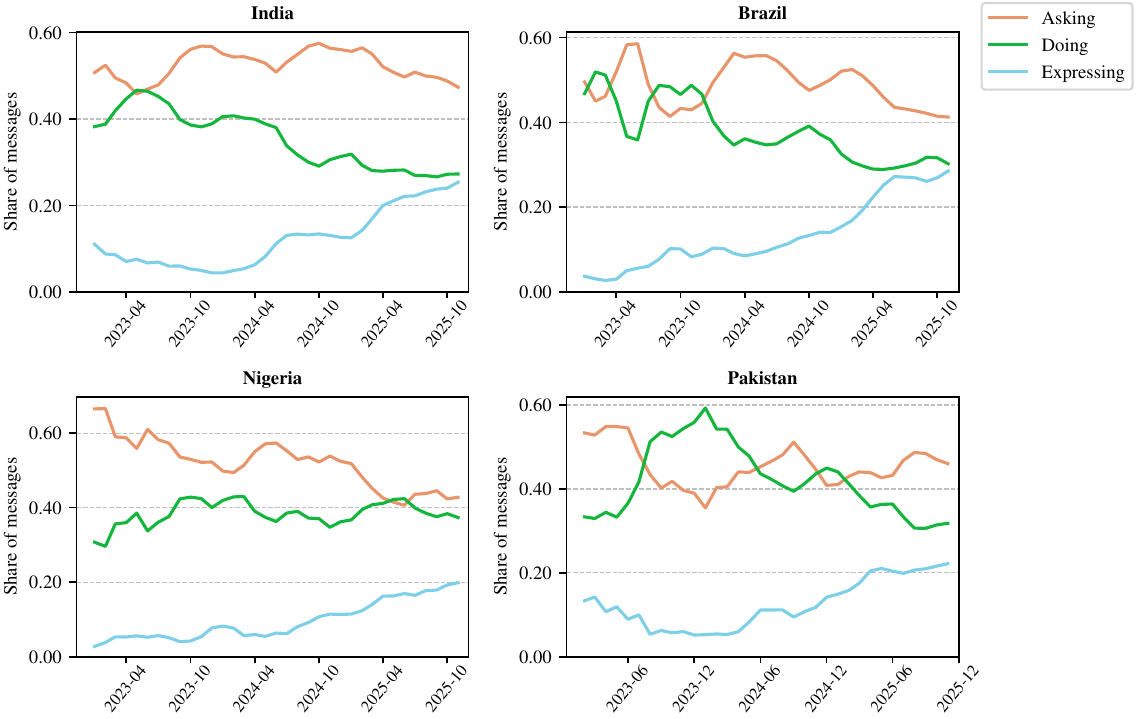}
    \caption{Temporal shift in \textit{Asking}/\textit{Doing}/\textit{Expressing} share (3-month rolling average) by country.}
    \label{fig:ask_do_express_time}
    \vspace{-\baselineskip}
\end{figure}

\textit{Asking} is the dominant intent throughout the observation window in every country (Appendix Figure~\ref{fig:ade_country}), \rev{as it is in OpenAI's global data \citep{chatterji2025chatgpt}, but its share has gradually declined while \textit{Doing} has remained roughly stable, rising only in Nigeria (Figure~\ref{fig:ask_do_express_time}).} \textit{Expressing} has grown steadily across all four countries and accounts for roughly a fifth of conversations in each by the end of our observation window, consistent with~\citet{karnam2026bowling} on increasingly socially-framed ChatGPT interactions and~\citet{fang2026aiwrapped} on reflective heavy-user behavior.

\rev{Within-user trends (Section~\ref{sec:within-user}) indicate that this growth is behavioral rather than compositional. Comparing each user's first three active months with their last three (449 users, median 20 months apart), the \textit{Expressing} share rises within users by 13.7~pp (95\% CI [12.0, 15.3]), in every country (India $+15.7$, Brazil $+16.2$, Nigeria $+10.5$, Pakistan $+9.6$~pp, all intervals excluding zero), and for 81\% of users individually, whereas \textit{Asking} falls by 10.6~pp and \textit{Doing} by 3.0~pp overall. \rev{First-use cohorts move in parallel through calendar time, so the rise is not an effect of later entrants. Appendix~\ref{app:extra-figs} gives the per-country changes, the cohort comparison, and the monthly within-user series.}}

\begin{figure}[t]
    \centering
    \includegraphics[width=0.5\textwidth]{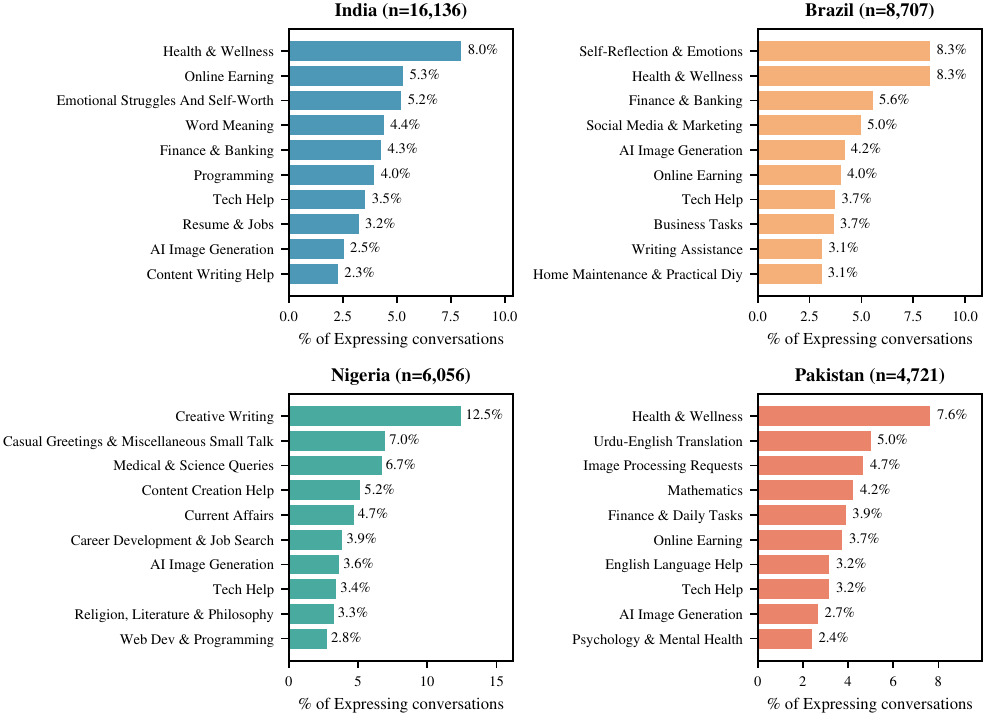}
    \caption{Top unsupervised clusters among \textit{Expressing} conversations, per country.}
    \label{fig:expressing_country_clusters}
    \vspace{-\baselineskip}
\end{figure}

\paragraph{Topics.}
Topical signatures are clean. Programming, content writing, resume drafting, and translation skew strongly toward \textit{Doing}, while current-affairs, health, finance, and ``how-to'' queries skew toward \textit{Asking}. \textit{Expressing} concentrates around self-reflection, emotional venting, religion, and personal/relationship messages \rev{(Appendix Figure~\ref{fig:ade_topics_coarse} for OpenAI topics, Figure~\ref{fig:expressing_country_clusters} for unsupervised topics).} Work-related conversations are disproportionately \textit{Doing} in every country (Appendix Figure~\ref{fig:ask_do_express_vs_work}). Men are slightly more likely than women to engage in \textit{Doing} while women lean marginally more towards \textit{Asking} and \textit{Expressing}, and the \textit{Doing} share grows monotonically across age bands (Appendix Figures \ref{fig:ade_demographics_gender}, \ref{fig:ade_demographics_age}). \rev{We read these demographic differences in intent as descriptive tendencies within this sample rather than as statistically supported differences in population.}

\paragraph{Language.}
Figure~\ref{fig:expressing_country_clusters} narrows in on \textit{Expressing} conversations only and lists the top unsupervised clusters in each country, showing what users actually express about. \rev{Two further measurements describe what this category contains. The first is linguistic. For each country we take the share of conversations in the dominant language (English for India, Nigeria and Pakistan, Portuguese for Brazil) and compare it across intents. \textit{Doing} conversations are conducted in the dominant language at substantially higher rates than \textit{Expressing} conversations in every country (88.6\% vs.\ 68.7\% in India; 96.2\% vs.\ 85.4\% in Nigeria; 85.2\% vs.\ 76.3\% in Brazil; 86.7\% vs.\ 73.9\% in Pakistan), with \textit{Asking} in between (Appendix Figure~\ref{fig:lang-ade}). Whether this reflects the topics of expressive conversations rather than their mode can be checked within the topic: across each country's 15 largest \textit{Expressing} clusters, the median gap to the \textit{Asking} and \textit{Doing} conversations of the same cluster is $+0.8$~pp, so most of the aggregate difference reflects which topics each mode covers. In the clusters where the language does change with the mode the shift is large, such as casual greetings in Nigeria (50.6\% vs.\ 90.2\%) and emotional struggles and self-worth in India (70.8\% vs.\ 80.7\%), while for others, such as health and wellness in India (68.2\% vs.\ 68.9\%), there is no gap at all (Appendix Table~\ref{tab:expressing_language}).}

\paragraph{What expressive conversations contain.}

\rev{The second is the qualitative reading (Section~\ref{sec:expressing-theme}). It shows two regularities. First, conversations consisting only of emotional expression were rare in the sample we read. Nearly all contained a request for information, advice, or a task. Second, the disclosure and the request are not separable.
Expressive conversations cluster around health concerns, relationships, academic pressure, religious questions, money, and requests for advice, and in each the disclosed context is what the request is about. A user in India in their early twenties describes a long-distance relationship at length and then asks whether the age gap is acceptable in India. A user lists symptoms before asking what a newly prescribed medicine does. A student in Pakistan, in an exam period, asks whether working only near deadlines is ``practically good or bad.'' A user in Brazil pastes a scriptural verse and asks for explanation, describes a religious role they are preparing for and how their health condition is making the role hard to fulfil. The question is no longer about the verse. A Nigerian user writes that they have paid a huge sum to an online seller who has gone silent and asks what to do.
In all of these, the disclosure is what makes the answer specific. We return to what this means in the Discussion.}

\section{Discussion}
\label{sec:discussion}

\rev{OpenAI has already reported, globally and for India, that most ChatGPT use is not work and that adoption is growing fastest in lower-income countries \citep{chatterji2025chatgpt, openai2026indiachatgpt}. Across the four countries we study, work conversations are roughly a fifth of total usage. Personal use is the majority in every country. What conversation-level data from these four countries adds is what those aggregates contain: which non-work uses, for whom, in which languages, and how the mode of use changes as people gain experience. We take the three research questions in turn, then ask what centering these users changes, and close with limitations. Each subsection ends with what remains open.}

\paragraph{RQ1: Beyond the workplace productivity frame.}
The economics of generative AI has focused on labor-market exposure and workplace productivity~\cite{eloundou2023gpts,brynjolfsson2023generative,noy2023experimental,dellacqua2023navigating}. Our results do not contradict that literature. Work conversations in our data often look like familiar productivity use cases: writing, programming, professional email, resumes, job applications, interview preparation, and content production. But work is only a minority of observed usage. Personal conversations account for 55--64\% of use in every country, and coursework is roughly as prevalent as work.

This distinction matters because many benefits of LLM use are not captured by a workplace productivity lens. A student getting private help with coursework, a user translating between Urdu and English, a patient trying to understand symptoms before deciding whether to seek care, or a young worker exploring online-earning strategies may all be extracting real value. But that value is not well described as making an employee faster at a job. Economists call this kind of non-market activity household production~\cite{becker1965theory}. The bulk of what we see in our conversations is exactly that. Health and wellness is among the most prevalent themes in India and Brazil and a top-5 theme in Pakistan, in countries where the ratio of physicians to population is a fraction of OECD levels. Translation is the top cluster in Pakistan.

\rev{The work we do observe is shaped by who the participants are. Clickworker recruits are gig workers and college students, and their work conversations (resumes, applications, client correspondence, content for short-video platforms) are the work of that population. An office-worker sample would show more work use and different work. The result that survives this caveat is that personal use is the majority in every age band. The coursework prevalence also deserves attention. Two readings are possible and neither is tested here: ChatGPT may serve as a private, low-friction source of academic help for users who face social, financial, or institutional barriers to asking elsewhere, or it may compensate for unequal access to offline educational resources. \citet{otis2024global} argue that generative AI's effect on inequality depends on whether it amplifies existing advantages or compensates for existing disadvantages. The pattern here makes that question measurable. Either way, education is a central domain of use in these countries rather than a spillover from productivity. What remains open is whether this non-market use improves outcomes: we observe that users treat ChatGPT as a first-line health or academic resource, not whether the answers they receive are accurate or whether the substitution is welfare-improving.}

\paragraph{RQ2: Local context shapes what the same tool becomes.}
The four countries in our study share broad patterns with global platform reports: writing, information seeking, practical guidance, and technical help are all common. But the unsupervised clusters show that the same global product is being used for different social functions in different settings. Health and wellness is among the most prevalent themes in India and Brazil and a top-5 theme in Pakistan. Religious queries are prominent in Nigeria and Pakistan but absent from India's and Brazil's top-15 lists. Urdu--English translation is the largest cluster in Pakistan. Online earning and YouTube monetization strategies are top clusters in India and Pakistan. Self-reflection is especially visible in Brazil.
What these patterns show is how users attach a general-purpose model to locally salient needs. For instance, the model becomes a first-pass health explainer in cases where access to medical advice is costly or uncertain, or a translator where multilingual navigation is part of daily life. The same interface is being pulled into different institutional and cultural gaps.

This is also where the taxonomy question matters. A fixed global taxonomy is useful for comparability, but it can flatten the most meaningful local uses. ``Practical Guidance'' does not tell us whether the user is asking about symptoms, religious practice, earning online, or a personal relationship. ``Translation'' does not tell us that Urdu--English translation is the dominant cluster in Pakistan. Global categories need to be paired with bottom-up discovery if we want to understand what adoption actually means in context. Future platform reporting would benefit from combining stable global categories with country-sensitive discovery pipelines. Neither alone is sufficient. The same logic applies to demographic inference. Name-based gender attribution is known to misclassify gender-ambiguous names (a particular concern for Indian names), and our self-reported labels reproduce the patterns OpenAI infers for India~\citep{openai2026indiachatgpt} while surfacing patterns (most strikingly the coursework difference above) that are easier to read at the individual-conversation level. \rev{Two things remain open. Whether the country differences reflect national contexts or the composition of each country's sample cannot be settled with convenience samples. It needs samples that are comparable across countries on age, gender, education, and occupation. And the demographic differences within countries are, with four exceptions, not individually reliable at this sample size, so the ``who'' of these uses is a direction rather than a finding.}

\paragraph{RQ3: Disclosure plus request.}

\rev{The intent results do not show a shift toward task delegation. In these countries \textit{Doing} is roughly flat and \textit{Asking} declines slowly. The mode that grows is \textit{Expressing}. Two explanations remain once composition is ruled out below. Users who learned that disclosing context produces more useful answers kept doing it, or the model, which became more conversational over the window, now elicits more disclosure. Our data cannot separate these two.}

\textit{Expressing} is the more distinctive signal. By the end of our observation window it accounts for roughly a fifth of conversations in every country and is growing steadily, a mode of use prior information tools could not host. Adopting it requires a different shift than \textit{Doing}. Users have to learn they can disclose personal context or vulnerability to a machine and expect a useful response, a trajectory aligned with \citet{karnam2026bowling} and \citet{fang2026aiwrapped}. \rev{The within-user analysis rules out that the change is due to composition of active users: we observe that the same users express more as their own histories progress ($+13.7$~pp from their first three active months to their last three, \textit{Expressing} share grows for 81\% of users individually).}

The qualitative reading shows the recurring pattern: users disclose personal context and then ask the model to do something with it, interpret a relationship, explain a medical worry, advise on a life decision, interpret a scriptural passage. This hybrid form, disclosure plus delegation, is what makes conversational AI different from search. \rev{A search engine can answer ``what does this medicine do''. It cannot answer the question after the user has attached symptoms, fear, or religious identity. Search trained users to minimize context. Chatbots train them to do the opposite, because more context produces more useful answers. A small number of conversations go further. Users name the assistant, refer back to earlier sessions, or scold it for a botched calculation, and for them the relationship has shifted past instrumental use. We report these as illustrations rather than as a measured category. The language pattern points the same way: \textit{Expressing} conversations are substantially less likely to be in the dominant language of each country, therefore more likely to be in user's native language, than \textit{Doing} conversations.  Task delegation often involves code, structured instructions, or professional templates that draw on English or Portuguese even in multilingual settings. Whether users also switch languages when the conversation turns personal, independently of topic, is what the within-theme comparison tests. Most of that difference turns out to be topic composition: across each country's 15 largest Expressing clusters, the median gap to the Asking and Doing conversations of the same cluster is only +0.8 pp. It stays large where the language changes with the mode, as in casual greetings in Nigeria (50.6\% vs. 90.2\%) and conversations about emotional struggles and self-worth in India (70.8\% vs. 80.7\%) (Appendix Table~\ref{tab:expressing_language}).}

The implications are double-edged. More context produces more relevant help, especially where users lack other resources. But the same context often involves health, mental health, finances, religion, relationships, or identity, and users are disclosing increasingly intimate material as they get more proficient with the tools. Safety systems that treat emotional support and task completion as separate domains will miss the overlap, because the risky cases are often both at once. The concern is sharpened commercially. Search engines trained users to be terse, which limited what advertisers could learn from a query. Chatbots train them to be expansive about exactly the material advertisers most want, just as the platforms hosting these conversations begin to monetize them. \rev{Because expressive conversations are disproportionately not in the country's dominant language, multilingual and culturally aware safety evaluation is central rather than peripheral.} And if users are turning to ChatGPT for companionship or emotional processing at scale, questions about parasocial attachment~\citep{turkle2017alone} and displacement of other forms of emotional support are real, and not visible from topic-level aggregates. \rev{What remains open is causal: whether disclosure improves the help users receive, whether the growth of expressive use reflects users or models, and what the disclosure habit does when it transfers to other systems.}

\paragraph{What centering these users changes.}
\rev{Centering non-Western users changes the understanding of a technology whose defaults were set elsewhere, and it changes the imagined user in three ways. The imagined user of the productivity literature is a salaried knowledge worker whose gains can be measured in output per hour. The typical user in our data is a student or gig worker for whom ChatGPT is a tutor, a translator, and a first-line health explainer, and whose gains are in household production that no employer records. The imagined user of safety evaluation writes in English and either seeks information or delegates a task. The typical expressive conversation here does both at once and is less often in the country's dominant language, so evaluations that test emotional-support and task-completion behavior separately, and in English, test a case these users rarely present. The imagined user of platform reports has a name-inferred gender. Ours self-report, which is how the coursework pattern becomes visible and how the name-inference problem for South Asian names can be checked. The point is not that these users are special cases. The defaults (English, work first, task and emotion kept apart) were set for someone else, and a general-purpose product in these markets is used by people the defaults did not anticipate. Three implications follow. Platform reporting should pair global categories with country-sensitive discovery. Safety and design evaluation should treat multilingual, disclosure-laden requests as the typical case rather than an edge case. And as chatbots move toward advertising, data protection for intimate disclosures matters most in exactly the markets where it is weakest.}

\paragraph{Limitations.}
\rev{Several limitations qualify our findings.
First, the sample. Participants were recruited through Clickworker and are digitally literate, largely English-proficient, and gig-economy-adjacent. They are younger and more male than the adult population in India and Pakistan and older and more female in Brazil, and the sample is not stratified on education or urban residence. Fewer of them hold salaried office jobs than the general population, so the work share is likely lower and differently composed than in an office-worker sample, and the prominence of online-earning clusters is plausibly amplified by the channel. All prevalence estimates are conditional on this population. The within-user analyses (Section~\ref{sec:within-user}) do not depend on it representing anyone but the participants themselves.
Second, the product. This is a study of ChatGPT use, not of AI use: people who reach AI through assistants embedded in WhatsApp or search engines, or through Gemini, do not appear here, and deliberate chatbot use is a narrower behavior than ambient exposure to AI-generated content. Our claims are about what happens when people open ChatGPT.
Third, measurement. The topic and intent constructs are defined by the platforms and implemented with the platform's own models (GPT-5-mini), so the instrument shares a vendor with the product. Validation shows substantial agreement for task purpose ($\kappa=0.82$) and lower agreement for intent ($\kappa=0.55$), driven by the \textit{Expressing} class, whose precision of 0.35 means that raw \textit{Expressing} shares are upper bounds.
Fourth, the sample of 1,252 users is small for subgroup comparisons: of the twenty-one demographic differences whose bootstrap intervals exclude zero, four survive correction. Fifth, our observation window coincides with rapid change in the model and its features, so user learning cannot be fully separated from changes in what the system elicits. Finally, we observe behavior but not outcomes: we can see that users ask health questions, but not whether the answers were accurate or helpful.}

\section{Conclusion}

\rev{We analyzed 202,590 ChatGPT conversations donated by 1,252 users in India, Nigeria, Brazil, and Pakistan, paired with self-reported age and gender: to our knowledge the first conversation-level, demographically grounded comparison of ChatGPT use across several Global South countries. Personal use is the majority in every country and coursework is about as common as work, so workplace productivity describes a minority of use. Unsupervised topic discovery shows the same product attached to different local needs (health, religion, translation, online earning, self-reflection) that a fixed taxonomy folds into generic categories. Over three years information seeking declined only modestly and task delegation did not grow, while expressive conversations, which attach a personal disclosure to a concrete request and are less often in the country's dominant language, grew to a fifth or more of use. Understanding what adoption means requires conversation-level, country-sensitive measurement alongside global aggregates. We release the classifier prompts, per-conversation labels, and aggregate tables, and will provide controlled access to the de-identified conversations for academic research.}

\section*{Generative AI use}
We acknowledge the use of generative AI tools, such as Gemini and Claude for ideation, help with framing, and writing assistance in parts of the paper. We also used Claude Code for data analysis. All the AI assisted writing and data analysis were thoroughly checked by the authors.

\bibliography{aaai25}

\clearpage
\appendix
\section{BERTopic Pipeline Details}
\label{app:bertopic-detail}

This appendix documents the full unsupervised topic-modeling pipeline summarized in Section~\ref{sec:methods}.

\textbf{Embedding.  } Each conversation is represented by a dense semantic embedding obtained from Google's \texttt{gemini-embedding-001} model, which produces 3{,}072-dimensional vectors. The embedding input is the first ten user--assistant turns of the conversation, serialized in the \texttt{[User]: ...$\backslash$n[Assistant]: ...} format, truncated at a per-message cap of 5{,}000 characters. Inputs exceeding the model's token limit are truncated by the API. Using the first ten turns rather than the whole conversation keeps per-conversation cost bounded and, in practice, captures the topic of the conversation well (users typically establish their task in the opening exchanges). All embedding vectors are L2-normalized before clustering.

\textbf{Initial clustering.  } For each country we fit MiniBatch $K$-means with $k=500$ on the embedding matrix. The deliberately high $k$ produces narrow, internally coherent micro-clusters (e.g.\ ``debugging Python code for web scraping'' rather than the generic ``programming''). Starting from an over-segmented solution lets the subsequent aggregation be driven by semantic similarity rather than forced early generalization.

\textbf{Hierarchical agglomeration.  } The 500 $K$-means centroids are then merged with agglomerative hierarchical clustering using cosine distance. The cutoff is selected per country by visual inspection of the dendrogram for natural breakpoints, targeting approximately 50 interpretable top-level topics, a number that balances cross-country comparability with sufficient granularity to reveal local patterns.

\textbf{Topic labeling and refinement.  } Each resulting cluster is labeled by summarization with \texttt{gpt-4o-mini}, using the top documents and most distinctive keywords in the cluster as context. To enhance label validity, we used Claude Sonnet 4.6 as a judge: clusters with ambiguous or low-confidence labels were either relabeled or flagged as non-homogeneous. Conversations in flagged clusters were then reassigned to defined topics by cosine-similarity scoring against cluster centroids, using a per-cluster $z$-score threshold of $1.5$ as the reassignment floor. The final pipeline yields \textbf{50}, \textbf{45}, \textbf{36}, and \textbf{53} topics for India, Nigeria, Brazil, and Pakistan respectively, covering 91--95\% of conversations per country. The remaining unassigned conversations (typically very short, off-topic, or multilingual/code-heavy) are excluded from BERTopic analyses but retained elsewhere.

\textbf{Cross-country theme grouping.  } For the ten-theme cross-country view in Section~\ref{sec:topics}, we group the per-country clusters into ten broad themes (Programming/Tech, Finance/Earning, Writing/Creative, Religion, Translation/Language, Health/Medical, Education/Academic, Job/Career, Content Creation, Current Affairs) by keyword matching on the cluster labels. This is a lossy aggregation (some cluster labels match multiple themes) but it makes the four countries comparable on a shared vocabulary while retaining the per-country cluster resolution for the cluster-level figures.

\section{Classifier validation}
\label{sec:classifier_validation}

To assess the reliability of the two LLM-based conversation-level classifiers - \textit{Work/Coursework/Personal} (Section~\ref{sec:work}) and \textit{Asking/Doing/Expressing} (Section~\ref{sec:ade}), we hand-labeled a stratified random sample of 50 conversations per country (194 valid items pooled across India, Brazil, Pakistan, and Nigeria after dropping ambiguous cases). One of the authors, blind to the model's prediction, assigned a gold label using the same definitions given to the classifier prompt. \rev{Conversations in Urdu and Portuguese were read in English translation, as in Section~\ref{sec:expressing-theme}.}
We then computed pooled accuracy, macro-F1, and Cohen's $\kappa$ against the model labels (Table~\ref{tab:val_summary}).

Both classifiers reach substantial pooled agreement (Table~\ref{tab:val_summary}). The \textit{Work/Coursework/Personal} classifier is reliable: pooled accuracy is 0.88 and $\kappa = 0.82$ (substantial agreement), with per-country accuracy ranging from 0.84 (Nigeria) to 0.94 (India). Per-country agreement for both classifiers is in Table~\ref{tab:val_by_country}. The \textit{Coursework} class is particularly clean (per-class F1 $\approx 0.95$). The dominant error mode is the model labeling some \textit{Personal} conversations as \textit{Work} (15/194 pooled cases), reflecting genuine overlap when personal queries borrow professional/work-style language.

For the \textit{Asking}/\textit{Doing}/\textit{Expressing} classifier (a categorical, single-label assignment per message, aggregated to conversation level by majority vote), the Asking and Doing classes are individually reliable ($\mathrm{F1} = 0.84$ and $0.77$), but Expressing has high recall (0.78) and low precision (0.35): the classifier frequently labels conversations as \textit{Expressing} when the gold label is \textit{Asking} or \textit{Doing}. This is consistent with the qualitative finding in Section~\ref{sec:ade}: very few \textit{Expressing} conversations in our corpus are purely affective, and most interleave a personal disclosure or emotional framing with a concrete request for information or task delegation. The classifier's recall on the human-gold Expressing class is therefore high. The precision drop reflects the same hybrid character, where conversations the annotators called \textit{Asking} or \textit{Doing} also contained enough self-disclosure for the model to predict \textit{Expressing}. We treat the classifier's Expressing prevalence as a noisy upper bound and report Expressing-conditioned results with that caveat.

\begin{table}[h]
  \centering
  \begin{tabular}{l c c c}
    \hline
    \textbf{Task} & \textbf{Acc.\ (95\% CI)} & \textbf{Macro-F1} & \textbf{$\kappa$} \\
    \hline
    W/C/P & 0.88 (0.83--0.92) & 0.88 & 0.82 \\
    A/D/E  & 0.81 (0.75--0.86) & 0.70 & 0.55 \\
    \hline
  \end{tabular}
  \caption{Pooled validation metrics against human gold labels on a stratified sample of $\sim$50 conversations per country. Both tasks reach substantial agreement. The residual error in Asking / Doing / Expressing is driven mainly by over-prediction of \textit{Expressing} (see text).}
  \label{tab:val_summary}
\end{table}

\begin{table}[h]
  \centering
  \begin{tabular}{l r r r}
    \hline
    \textbf{Country} & \textbf{$n$} & \textbf{W/C/P $\kappa$} & \textbf{A/D/E $\kappa$} \\
    \hline
    India    & 47  & 0.90 & 0.51 \\
    Brazil   & 50  & 0.79 & 0.72 \\
    Pakistan & 48  & 0.84 & 0.53 \\
    Nigeria  & 49  & 0.76 & 0.44 \\
    \hline
    \textbf{Pooled} & \textbf{194} & \textbf{0.82} & \textbf{0.55} \\
    \hline
  \end{tabular}
  \caption{Per-country Cohen's $\kappa$ for each classifier.}
  \label{tab:val_by_country}
\end{table}

\section{Statistical Significance Results}
\label{app:stats-sig}

\begin{table}[t]
\centering

\small
\begin{tabular}{@{}llrcc@{}}
\hline
\textbf{Country} & \textbf{Purpose} & \textbf{M $-$ F} & \textbf{95\% CI (pp)} & \textbf{$q$} \\
\hline
India    & Coursework & $-8.18$  & $[-14.97,\;-2.10]$ & .043* \\
India    & Work       & $+3.76$  & $[-0.71,\;8.25]$   & .150 \\
India    & Personal   & $+4.41$  & $[-3.14,\;12.20]$  & .263 \\
\hline
Nigeria  & Personal   & $+8.53$  & $[-0.64,\;17.39]$  & .201 \\
Nigeria  & Work       & $-4.05$  & $[-9.82,\;1.96]$   & .268 \\
Nigeria  & Coursework & $-4.49$  & $[-11.97,\;3.58]$  & .273 \\
\hline
Brazil   & Coursework & $+2.56$  & $[-3.58,\;9.49]$   & .993 \\
Brazil   & Personal   & $-2.59$  & $[-13.32,\;9.24]$  & .993 \\
Brazil   & Work       & $+0.03$  & $[-10.32,\;9.28]$  & .994 \\
\hline
Pakistan & Coursework & $-11.91$ & $[-25.77,\;3.41]$  & .207 \\
Pakistan & Personal   & $+10.44$ & $[-3.68,\;23.40]$  & .207 \\
Pakistan & Work       & $+1.47$  & $[-6.45,\;8.92]$   & .706 \\
\hline
\end{tabular}
\caption{\rev{Gender differences in task purpose by country (Section~\ref{sec:work}). Each row is the difference between men's and women's share of conversations in that category, in percentage points, with a 95\% user-clustered bootstrap interval and a Benjamini--Hochberg $q$-value across the three categories within a country. Negative values indicate a higher share among women. The asterisk (*) marks the only difference that is significant after correction.}}
\label{tab:work_gender_results}
\end{table}

\begin{table}[t]
\centering

\begin{tabular}{llll}
\hline
\textbf{Country} & \textbf{Topic} & \textbf{Comp.} & \textbf{95\% CI} \\
\hline
India & Tutoring/Teaching & F $>$ M & [-6.08, -0.71] \\
Brazil & Programming & M $>$ F & [1.49, 7.47] \\
Brazil & Personal Writing & F $>$ M & [-8.57, -1.58] \\
Nigeria & Cooking/Recipes & F $>$ M & [-1.31, -0.25] \\
Nigeria & Health/Self-Care & F $>$ M & [-4.49, -0.50] \\
Pakistan & Products & M $>$ F & [0.26, 1.81] \\
Pakistan & Personal Reflection & M $>$ F & [0.04, 1.31] \\
\hline
\end{tabular}
\caption{Gender-topic differences whose 95\% user-clustered bootstrap confidence intervals exclude zero. Positive intervals indicate higher prevalence among men and negative intervals higher prevalence among women. None remain significant after Benjamini--Hochberg correction.}
\label{tab:taxonomy_gender_results}
\end{table}

\begin{table*}[t]
\centering

\begin{tabular}{@{}lllll@{}}
\hline
\textbf{Group} & \textbf{Country} & \textbf{Topic} & \textbf{95\% CI (pp)} & \textbf{$q$} \\
\hline Gender (M $>$ F) & Brazil & Programming/Tech & $[2.84,\;11.34]$ & .020 \\
Gender (F $>$ M) & Brazil & Job/Career & $[1.03,\;7.98]$ & .077 \\
Gender (F $>$ M) & Nigeria & Health/Medical & $[2.13,\;11.28]$ & .064 \\
Gender (F $>$ M) & Nigeria & Job/Career & $[0.69,\;6.41]$ & .064 \\
Gender (M $>$ F) & Nigeria & Finance/Earning & $[0.34,\;4.42]$ & .104 \\
Gender (M $>$ F) & Pakistan & Current Affairs & $[0.29,\;2.87]$ & .138 \\
Gender (M $>$ F) & Pakistan & Religion & $[0.32,\;4.83]$ & .224 \\
\hline
Age (18--25 $>$ 36+) & India & Education/Academic & $[1.49,\;9.12]$ & .050 \\
Age (18--25 $>$ 36+) & India & Programming/Tech & $[2.95,\;20.30]$ & .050 \\
Age (36+ $>$ 18--25) & India & Finance/Earning & $[3.04,\;23.38]$ & .050 \\
Age (18--25 $>$ 36+) & Brazil & Education/Academic & $[8.87,\;22.40]$ & .003* \\
Age (18--25 $>$ 36+) & Nigeria & Education/Academic & $[2.71,\;14.35]$ & .117 \\
Age (18--25 $>$ 36+) & Pakistan & Programming/Tech & $[10.16,\;29.02]$ & .006* \\
\hline
\end{tabular}
\caption{Demographic-topic differences whose 95\% user-clustered bootstrap confidence intervals exclude zero. Asterisks (*) denote differences that remain significant after Benjamini--Hochberg correction ($q<.05$). Positive intervals indicate higher prevalence for the group shown in the comparison column.} \label{tab:demographic_topic_results}
\end{table*}

\begin{table}[t]
\centering

\setlength{\tabcolsep}{3.5pt}
\begin{tabular}{@{}llrrr@{}}
\hline
\textbf{Country} & \textbf{Cluster} & \textbf{Expr.} & \textbf{A+D} & \textbf{Diff.} \\
\hline
India    & Word meaning                    & 50.4 & 72.0 & $+21.6$ \\
Nigeria  & Casual greetings, small talk    & 50.6 & 90.2 & $+39.6$ \\
Pakistan & Urdu--English translation       & 57.7 & 64.8 & $+7.1$ \\
Brazil   & Portuguese--English translation & 60.2 & 69.2 & $+9.0$ \\
India    & Current affairs                 & 64.7 & 67.9 & $+3.2$ \\
Nigeria  & English grammar \& language     & 66.3 & 89.3 & $+23.0$ \\
India    & Health \& wellness              & 68.2 & 68.9 & $+0.7$ \\
India    & Online earning                  & 70.3 & 73.3 & $+3.0$ \\
India    & Emotional struggles, self-worth & 70.8 & 80.7 & $+9.9$ \\
Pakistan & Finance \& daily tasks          & 73.1 & 74.7 & $+1.6$ \\
\hline
\end{tabular}
\caption{\rev{The ten leading \textit{Expressing} clusters with the lowest share of conversations in the country's dominant language (Section~\ref{sec:ade}). \textit{Expr.} is that share among a cluster's \textit{Expressing} conversations, \textit{A+D} among its \textit{Asking} and \textit{Doing} conversations in the same country, and \textit{Diff.} their difference in percentage points. Candidates are each country's 15 largest clusters by \textit{Expressing} volume (at least 96 such conversations each). Across those 60 clusters the median difference is $+0.8$~pp.}}
\label{tab:expressing_language}
\end{table}

\section{Ethics and Data Protection}
\label{app:ethics}

\rev{\textbf{Consent and recruitment.} Participants were recruited on Clickworker with a task description stating that the study collects complete ChatGPT conversation histories for research on how people use conversational AI, that names and contact details are removed on the participant's own device before upload, that raw conversations are never published, and that participation can be withdrawn. Participants gave consent on the study website before uploading and confirmed that the exported account was their own. The protocol was approved by our institutional review board. }

\rev{\textbf{Compensation.} Participants received a fixed payment per completed upload, set in each country at or above the local minimum hourly wage for the expected duration of the task.}

\rev{\textbf{De-identification.} The ChatGPT export is processed in the participant's browser before anything is transmitted. The client-side script removes personal names, e-mail addresses, and other direct identifiers from the export, including the account profile, before upload. The script cannot remove everything that could identify a person: participants describe life events, workplaces, and other people, and no automatic method removes those reliably. For this reason we treat the corpus as sensitive even after scrubbing, and the release policy below follows from it.}

\rev{\textbf{Storage and access.} Uploaded exports are stored on access-controlled institutional storage, accessible only to the research team, and are not shared with the crowd-work platform or with any AI vendor except as inputs to the classification and embedding API calls described in Section~\ref{sec:methods}, which are made under the vendors' API terms, under which submitted data are not used to train their models.}

\rev{\textbf{Release policy.} We considered a WildChat-style public release \citep{zhao2024wildchat} and decided against it for this dataset. WildChat conversations are single sessions from anonymous users. Ours are multi-year personal histories tied to age, gender, and country, in which the combination of disclosed life events can identify a person even with names removed, and participants consented to research use rather than to publication. We therefore release, at the link in the paper, the per-conversation labels (OpenAI topic, unsupervised cluster, intent, task purpose, detected language, and month) joined to participant age, gender, and country, together with the classifier prompts, the analysis code, and the aggregate tables behind every figure. This supports re-analysis of every quantitative result in the paper. The de-identified conversation text will not be released publicly. We will provide partial academic access to it: researchers at academic institutions may apply for access to a defined subset for a stated research purpose under a data-use agreement that prohibits re-identification, redistribution, and commercial use, with the agreement reviewed by our institution. A datasheet \citep{gebru2021datasheets} will accompany the released labels.}

\rev{\textbf{Examples in the paper.} The conversation examples quoted in Sections~\ref{sec:ade} and~\ref{sec:discussion} are paraphrased, stripped of exact ages, amounts, drug names, place names, and combinations of occupation and condition, and chosen so that no example rests on a rare combination of attributes.}

\section{Research Questions and Methods}
\label{app:rq-map}

\rev{Table~\ref{tab:rq_methods} maps each research question to the measurements that answer it, how each measurement is validated, and where the results appear. It is reproduced here rather than in Section~\ref{sec:methods} for space.}

\begin{table*}[t]
  \centering

  \small
  \begin{tabular}{p{2.4cm} p{5.4cm} p{4.6cm} p{2.6cm}}
    \hline
    \textbf{Question} & \textbf{Measurement} & \textbf{Validation} & \textbf{Results} \\
    \hline
    RQ1 Purpose & (iv) task-purpose classifier; unsupervised clusters within each purpose & 194 hand-labelled conversations, $\kappa=0.82$ & Section~\ref{sec:work} \\
    RQ2 Topics & (i) OpenAI 24-category taxonomy; (ii) unsupervised topic discovery per country & Comparison with OpenAI's published shares; judged and manually reviewed cluster labels & Section~\ref{sec:topics} \\
    RQ3 Manner of use & (iii) intent classifier; language detection; within-user trends; thematic analysis of expressive conversations & $\kappa=0.55$; \textit{Expressing} precision 0.35 & Section~\ref{sec:ade} \\
    Who (within each RQ) & Self-reported age and gender; user-clustered bootstrap with Benjamini--Hochberg correction & Only differences with intervals excluding zero are reported; those surviving correction are marked & Sections~\ref{sec:work}--\ref{sec:ade} \\
    \hline
  \end{tabular}
  \caption{Research questions, the measurements that answer them, how each measurement is validated, and where the results appear.}
  \label{tab:rq_methods}
\end{table*}

\section{Additional Figures}
\label{app:extra-figs}

This appendix collects the per-country, per-demographic, and per-topic breakdowns that supplement the body figures. The figures are organized by the three \rev{research questions} of Section~\ref{sec:results}: \emph{purpose} (work/coursework/personal), \emph{topics} (supervised and unsupervised), and \emph{intent} (Asking/Doing/Expressing). Each figure is referenced from the corresponding results subsection. We briefly describe what each shows below.

\subsection*{Purpose: per-task topics and trends}

\begin{figure*}[t]
  \centering
  \includegraphics[width=\textwidth]{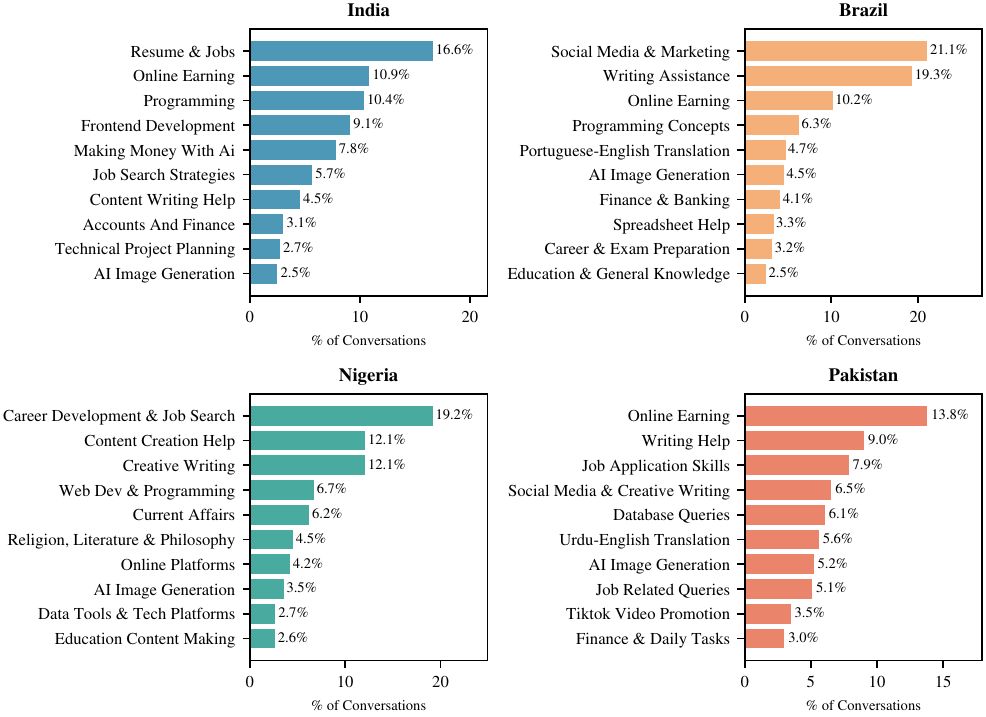}
  \caption{Top 10 unsupervised topic clusters for work-related conversations, by country (Section~\ref{sec:work}).}
  \label{fig:work-topics}
\end{figure*}

\begin{figure*}[t]
  \centering
  \includegraphics[width=\textwidth]{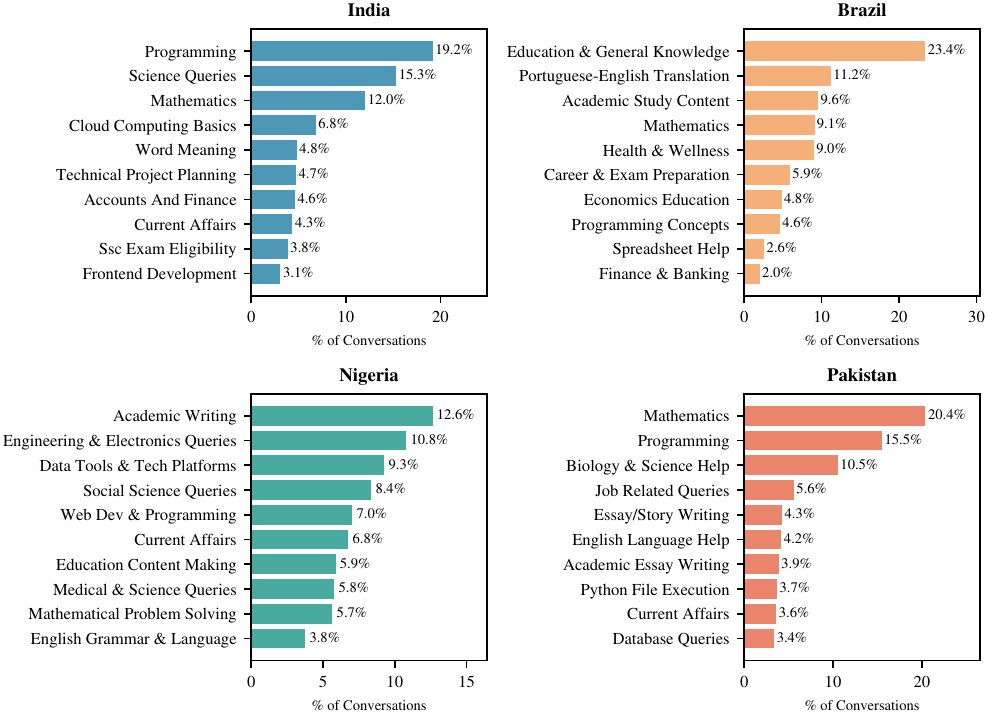}
  \caption{Top 10 unsupervised topic clusters for coursework conversations, by country (Section~\ref{sec:work}).}
  \label{fig:coursework-topics}
\end{figure*}

\begin{figure*}[t]
  \centering
  \includegraphics[width=0.7\textwidth]{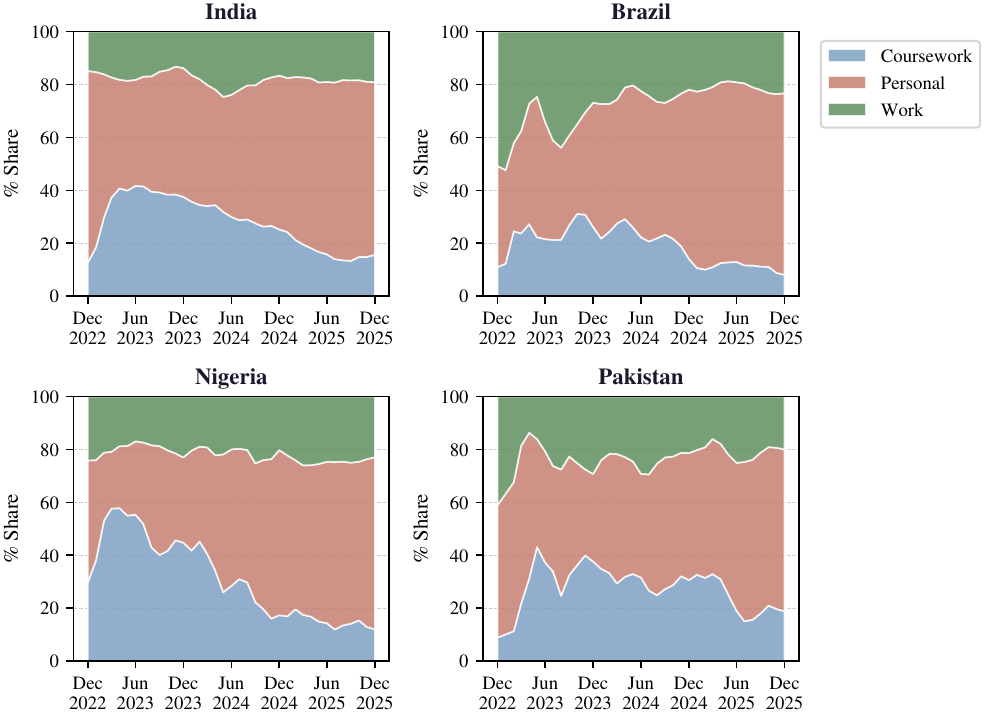}
  \caption{Temporal share of work, coursework, and personal conversations (3-month rolling average), by country (Section~\ref{sec:work}).}
  \label{fig:work_label_trends}
\end{figure*}

\begin{figure*}[t]
  \centering
  \includegraphics[width=\textwidth]{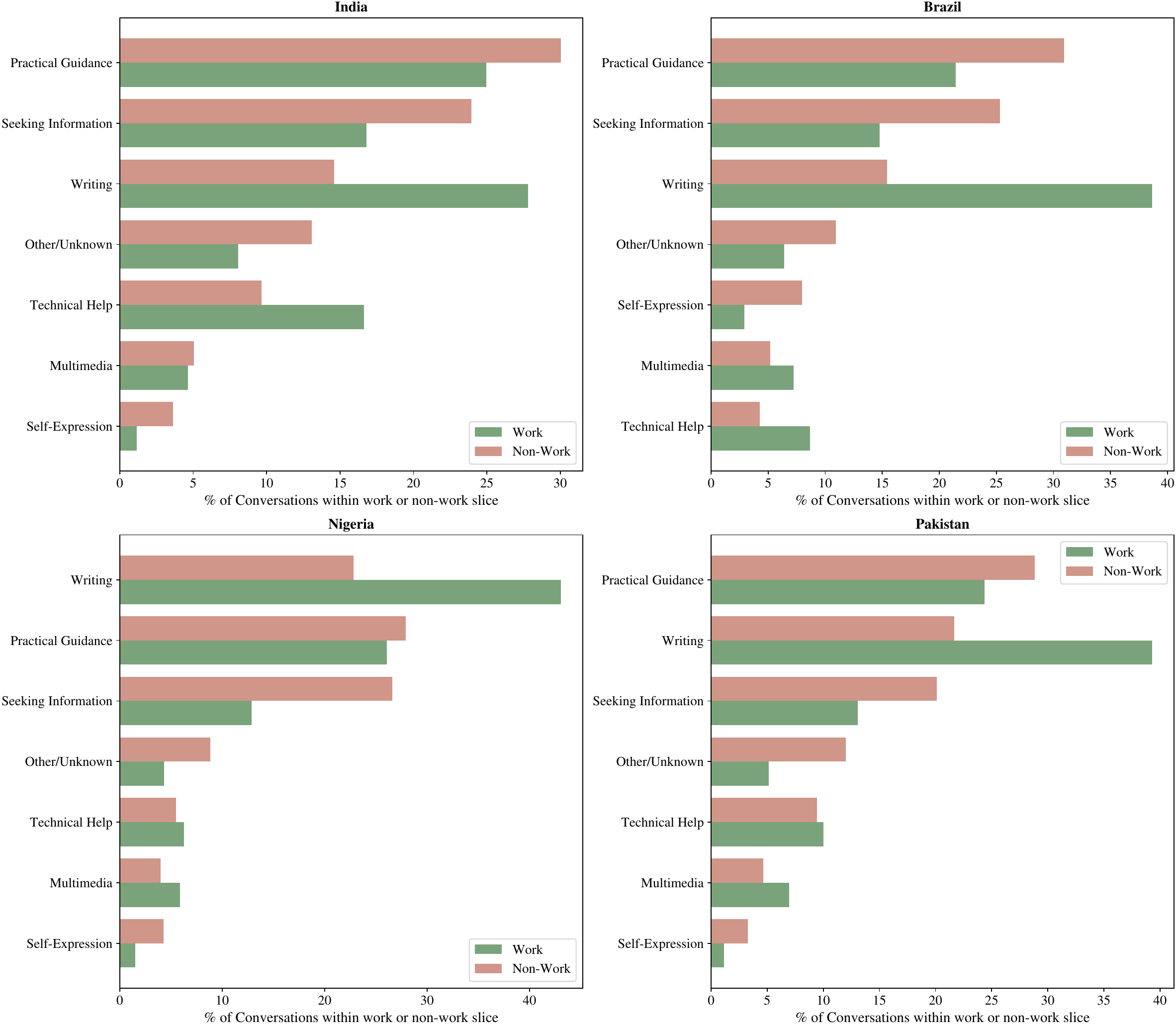}
  \caption{Share of coarse topics for work-related vs.\ non-work conversations, by country (Section~\ref{sec:work}).}
  \label{fig:openai-topic-work}
\end{figure*}

\rev{Figures~\ref{fig:work-topics} and~\ref{fig:coursework-topics} resolve the unsupervised BERTopic clusters under the work and coursework labels, showing the top-10 clusters per country. The personal-conversation counterpart is Figure~\ref{fig:personal-topics} in the body.} Figure~\ref{fig:work_label_trends} shows the temporal share of the three task-purpose labels (3-month rolling average), per country. Figure~\ref{fig:openai-topic-work} compares the OpenAI 24-category coarse topic distribution for work-related versus non-work conversations and shows the writing-heavy character of work use.

\begin{figure*}[t]
  \centering
  \includegraphics[width=0.70\textwidth]{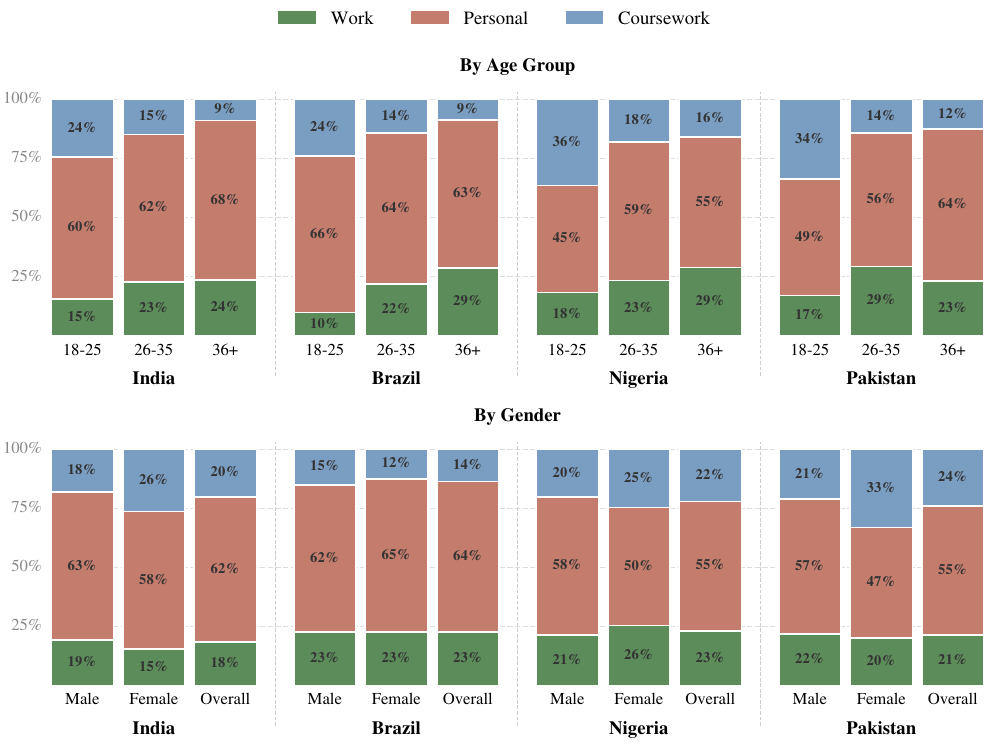}
  \caption{Distribution of task purpose (work, coursework, personal) by age group (top) and gender (bottom), by country.}

  \label{fig:work-personal}
  \vspace{-\baselineskip}
\end{figure*}

\FloatBarrier
\subsection*{Topics: full distributions and demographic conditioning}

\begin{figure*}[t]
  \centering
  \includegraphics[width=\textwidth]{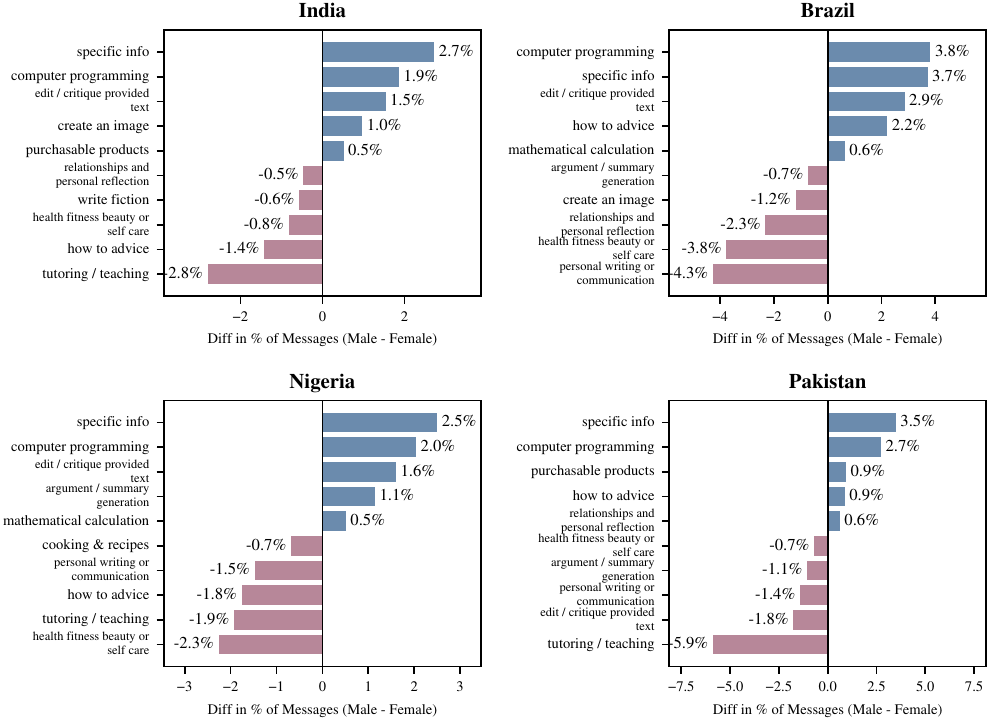}
  \caption{Gender topic divergence: difference in topic share (Male $-$ Female) per country under the OpenAI taxonomy. Blue indicates male over-representation and pink female over-representation (Section~\ref{sec:topics}). {The figure presents the five largest differences in each direction, selected by effect magnitude rather than statistical significance.}}
  \label{fig:gender-topics}
\end{figure*}

\begin{figure*}[t]
\includegraphics[width=0.9\textwidth]{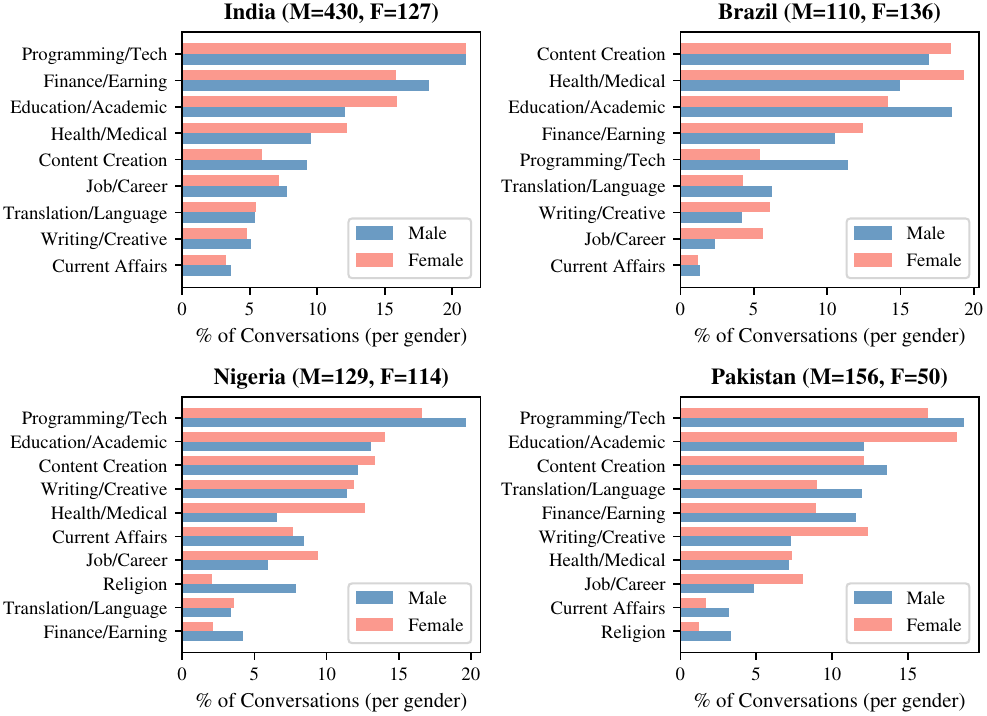}
\caption{Unsupervised themes by gender across countries (Section~\ref{sec:topics}). Each bar shows the share of a gender's conversations assigned to a theme. Male over-representation in programming, finance, and religion, and female over-representation in health, content creation, and education.}
\label{fig:gender_bertopic}
\end{figure*}

\begin{figure*}[t]
\includegraphics[width=0.9\linewidth]{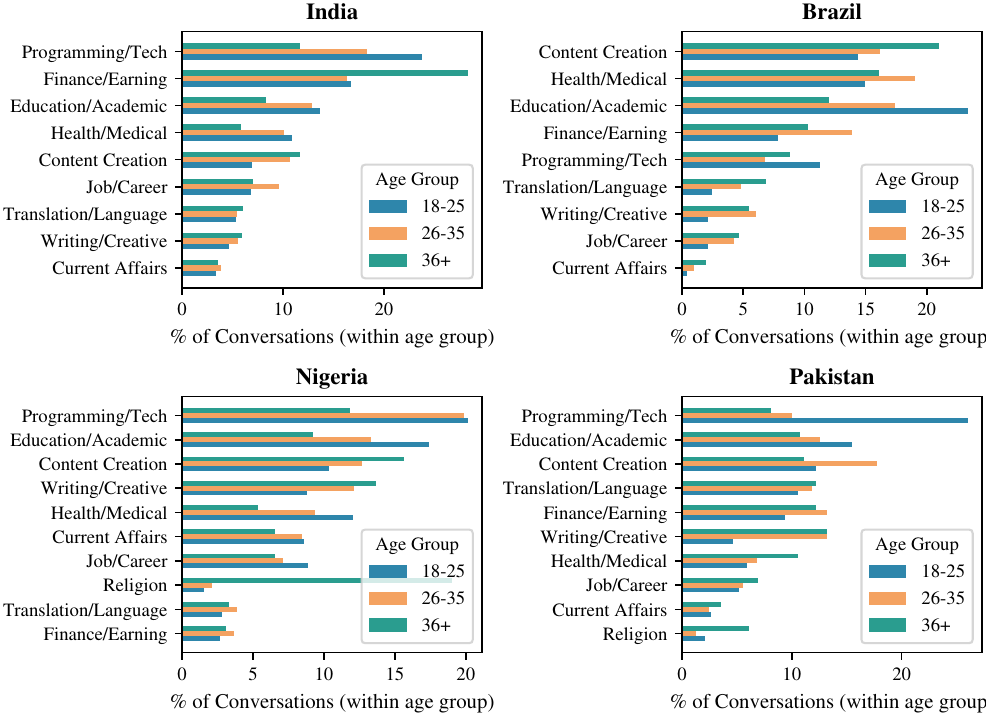}
\caption{Unsupervised themes by age group across countries (Section~\ref{sec:topics}). Younger users concentrate on education and programming. Older users shift toward finance, content creation, and religion.}
\label{fig:age_bertopic}
\end{figure*}

\begin{figure*}[t]
  \centering
  \includegraphics[width=0.8\textwidth]{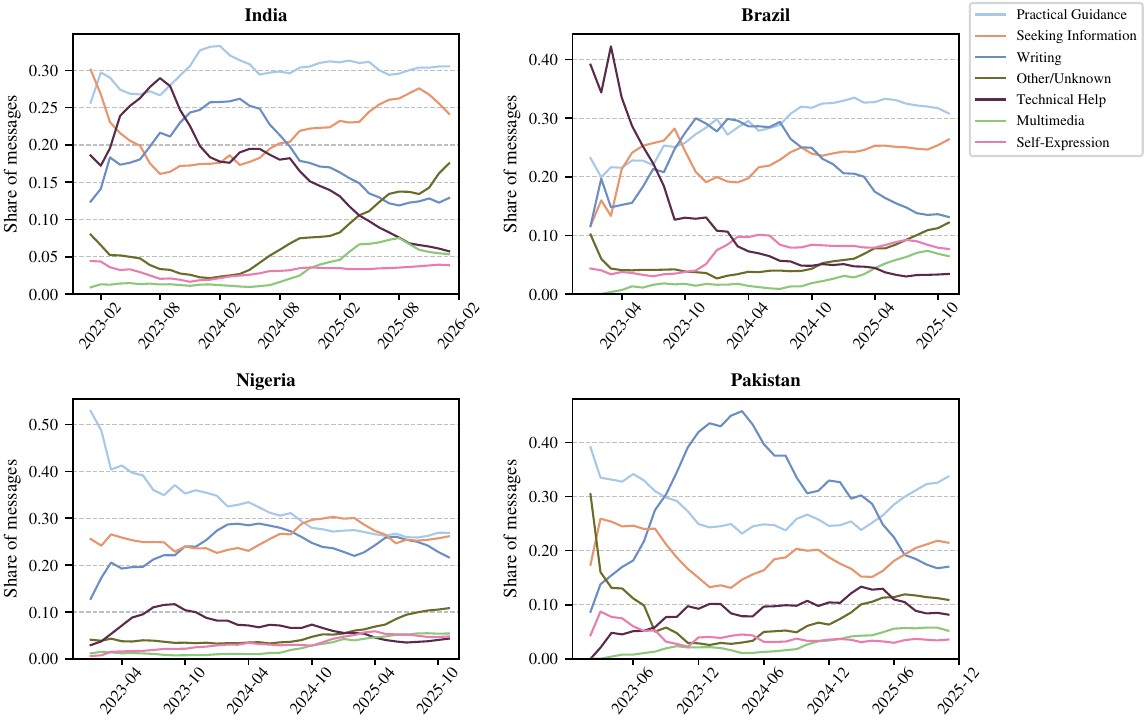}
  \caption{Share of coarse topics over time, non-work conversations only (Section~\ref{sec:topics}).}
  \label{fig:topic-trends-openai-lines}
\end{figure*}

Figure~\ref{fig:gender-topics} shows the gender-conditional topic divergence under the supervised OpenAI taxonomy (Male minus Female share per topic) for each country{. The differences whose bootstrap intervals exclude zero are listed in Table~\ref{tab:taxonomy_gender_results}, none of which survives correction}. \rev{Figure~\ref{fig:unsup_topics} in the body shows the top-15 unsupervised topic clusters per country, the granular view that the ten-theme heatmap (Figure~\ref{fig:theme_heatmap}) aggregates.} Figures~\ref{fig:gender_bertopic} and~\ref{fig:age_bertopic} show how the ten unsupervised themes break down by gender and age group, respectively. Figure~\ref{fig:topic-trends-openai-lines} shows the share of OpenAI coarse topics over time for non-work conversations, illustrating the early dominance of writing and technical help and the later rise of practical guidance and information seeking.

\FloatBarrier
\subsection*{Intent: demographic and topical conditioning}

\begin{figure*}[t]
    \centering
    \includegraphics[width=0.75\textwidth]{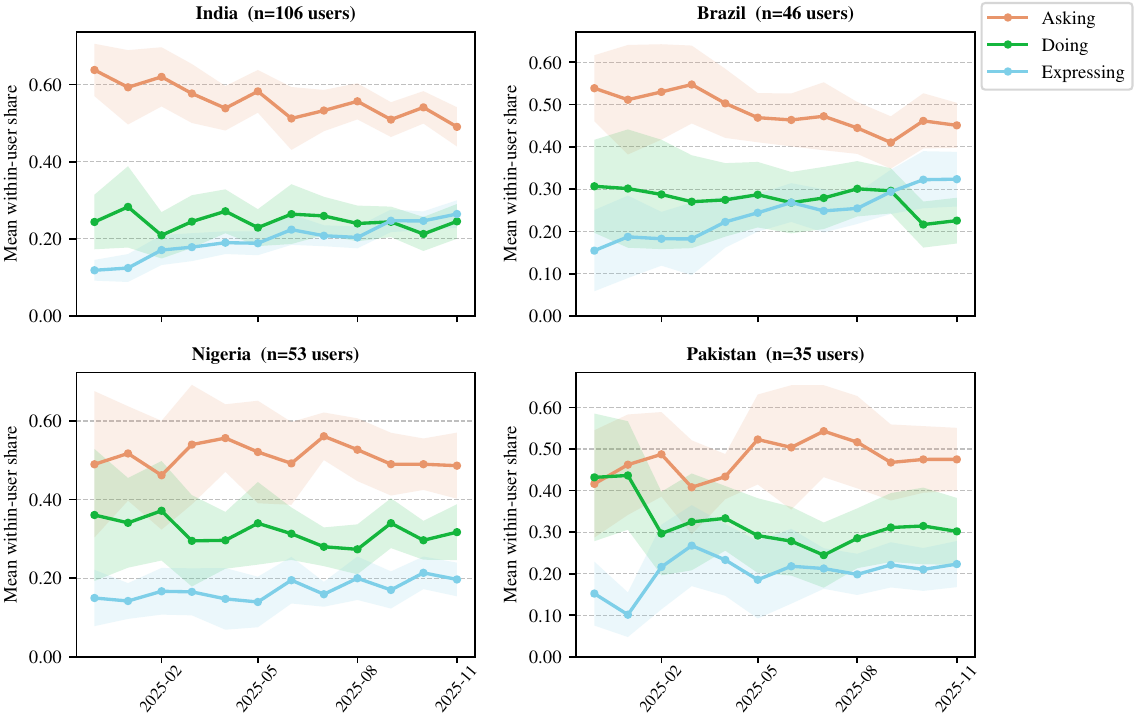}
    \caption{\rev{Within-user \textit{Asking}/\textit{Doing}/\textit{Expressing} shares by month, for the users with more than 30 conversations in a given month, over December 2024--November 2025. Each point is the mean of users' own monthly shares, so every user counts once. Bands are 95\% intervals across users.}}
    \label{fig:ade_within_user}
\end{figure*}

\rev{Within-user intent trends (Section~\ref{sec:ade}) hold in each country taken separately: between a user's first three active months and their last three, the \textit{Expressing} share rises by 15.7~pp in India, 16.2~pp in Brazil, 10.5~pp in Nigeria, and 9.6~pp in Pakistan, all with intervals excluding zero. Splitting users by the year of their first conversation gives curves that move together through calendar time rather than at different levels: for \textit{Expressing}, the quarterly curves of users who began in 2022--23 and in 2024 correlate at $r=0.98$, with a mean level gap of 2.7~pp across the eight quarters in which both are observed. Figure~\ref{fig:ade_within_user} shows the monthly within-user series for the most active users.}

\begin{figure}[t]
  \centering
  \includegraphics[width=\columnwidth]{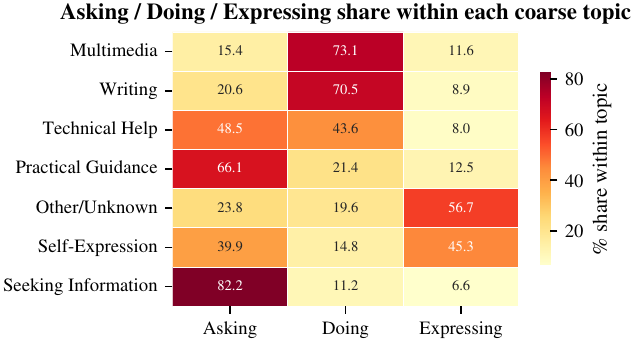}
  \caption{Share of \textit{Asking}/\textit{Doing}/\textit{Expressing} within each coarse topic, aggregated across all four countries, according to the OpenAI taxonomy,  (Section~\ref{sec:ade}).}
  \label{fig:ade_topics_coarse}
\end{figure}

\begin{figure*}[t]
  \centering
  \includegraphics[width=0.8\textwidth]{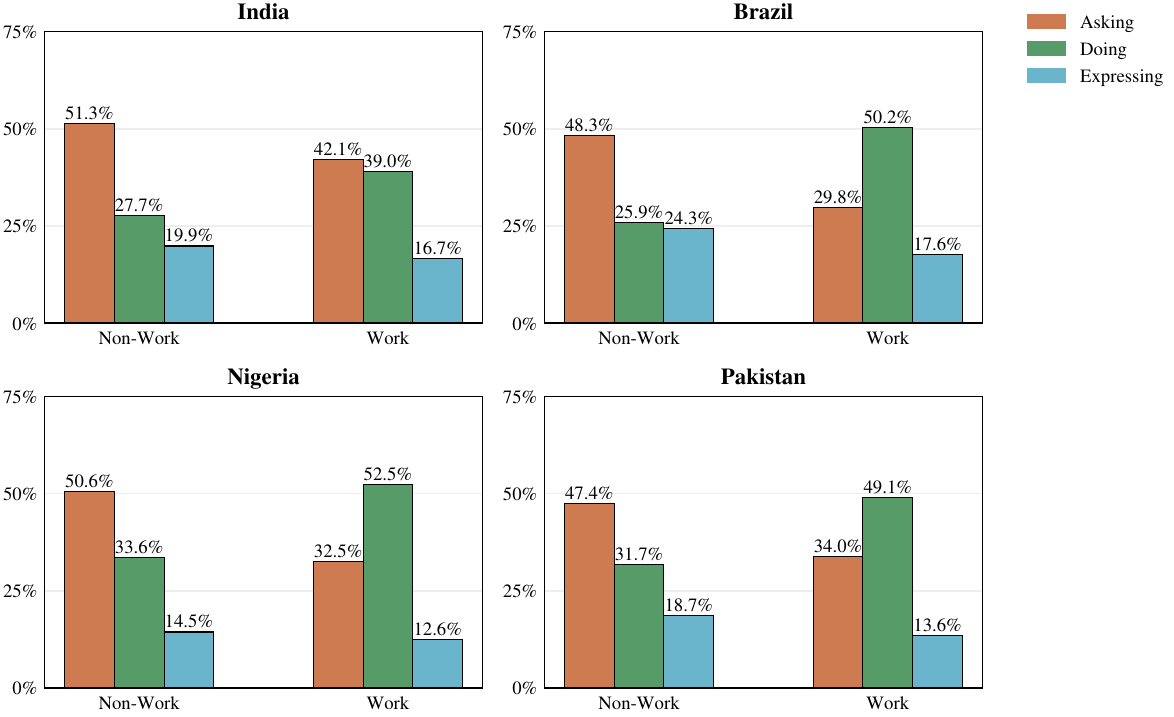}
  \caption{Share of \textit{Asking}/\textit{Doing}/\textit{Expressing} conversations by task purpose (work vs.\ non-work) and country (Section~\ref{sec:ade}).}
  \label{fig:ask_do_express_vs_work}
\end{figure*}

\begin{figure*}[t]
  \centering
  \includegraphics[width=\textwidth]{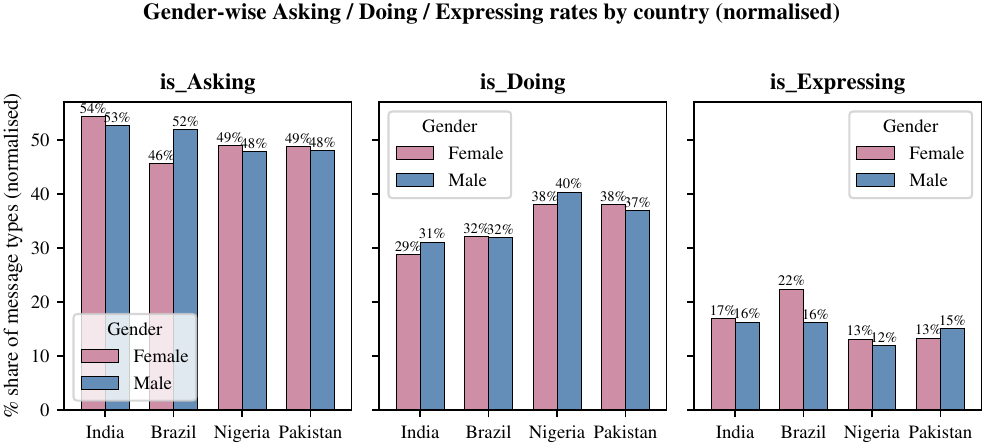}
  \caption{Normalized \textit{Asking}/\textit{Doing}/\textit{Expressing} share by gender, per country (Section~\ref{sec:ade}).}
  \label{fig:ade_demographics_gender}
\end{figure*}

\begin{figure*}[t]
  \centering
  \includegraphics[width=\textwidth]{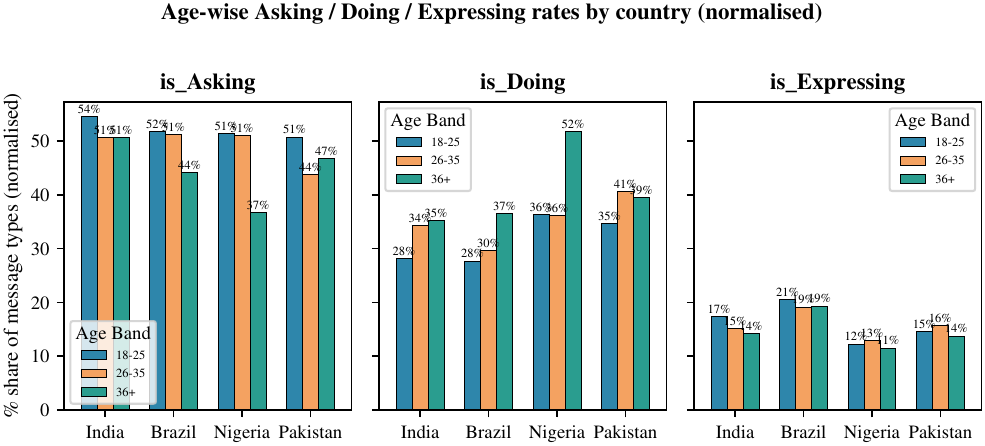}
  \caption{Normalized \textit{Asking}/\textit{Doing}/\textit{Expressing} share by age group, per country (Section~\ref{sec:ade}).}
  \label{fig:ade_demographics_age}
\end{figure*}

\begin{figure*}[t]
  \centering
  \includegraphics[width=\textwidth]{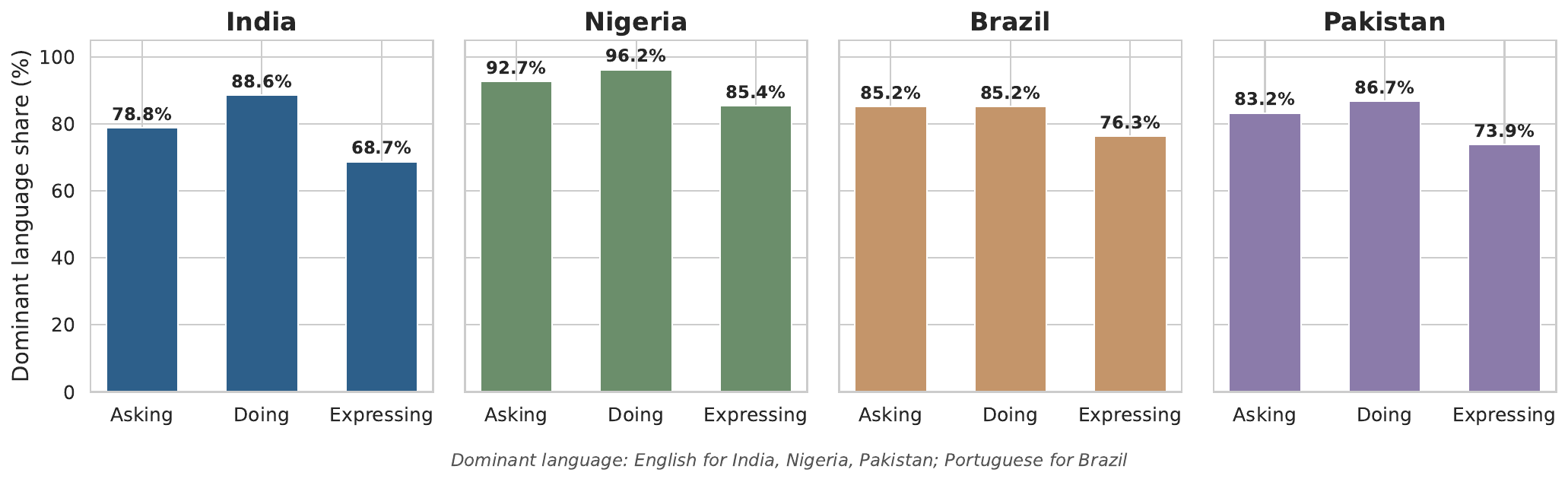}
  \caption{Dominant-language share by conversational intent (\textit{Asking}, \textit{Doing}, \textit{Expressing}) and country. Expressing has the lowest dominant-language share in every country (Section~\ref{sec:ade}).}
  \label{fig:lang-ade}
\end{figure*}

\begin{figure*}[t]
    \centering
    \includegraphics[width=0.7\textwidth]{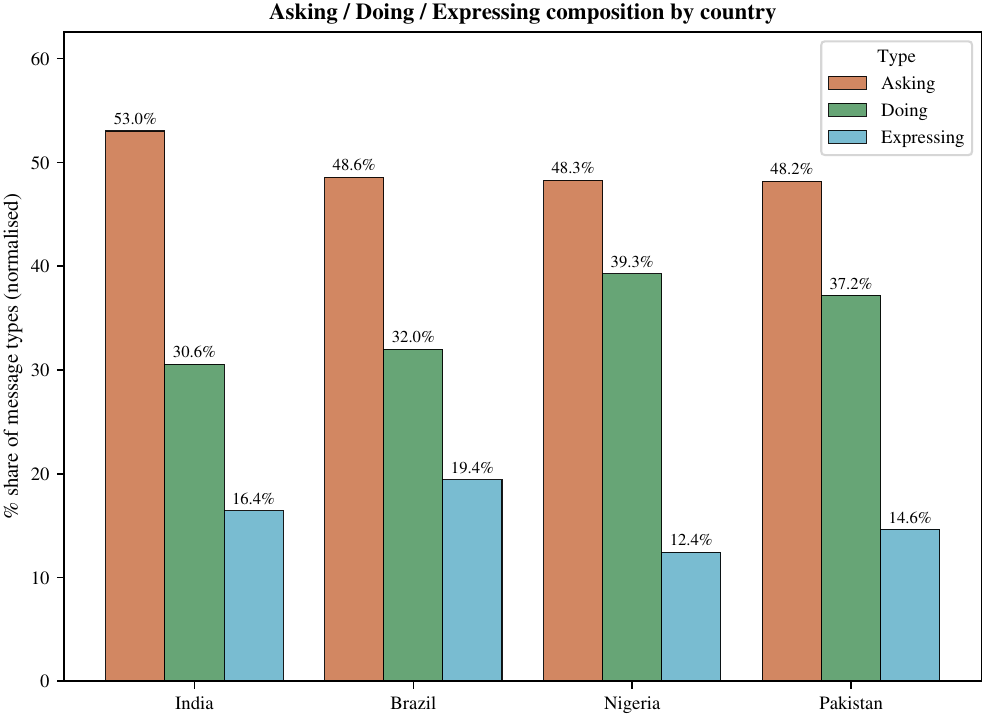}
    \caption{Normalized composition of \textit{Asking}/\textit{Doing}/\textit{Expressing} conversations by country (Section~\ref{sec:ade}).}
    \label{fig:ade_country}
\end{figure*}

Figure~\ref{fig:ade_topics_coarse} shows how \textit{Asking}, \textit{Doing}, and \textit{Expressing} are distributed within each OpenAI coarse topic, making the topic--intent association concrete (programming and translation skew Doing; health and how-to skew Asking; self-expression and personal communication skew Expressing). Figure~\ref{fig:ask_do_express_vs_work} contrasts the ADE composition for work versus non-work conversations. Figures~\ref{fig:ade_demographics_age} and~\ref{fig:ade_demographics_gender} show the ADE composition broken out by age group and gender respectively. Figure~\ref{fig:lang-ade} shows the dominant-language share by intent (Expressing has the lowest dominant-language share in every country). Figure~\ref{fig:ade_country} shows the overall ADE composition per country.

\end{document}